\documentclass{iopjournal}
\usepackage{amsmath}
\usepackage{subcaption}
\usepackage{algorithm}
\usepackage{algpseudocode}
\usepackage{amssymb}
\usepackage{float}

\usepackage{pdfpages}

\begin{document}

\articletype{Article type} 

\title{Learning and Generating Time Trajectories of Multidimensional Stochastic Processes}






\author{Zihan Xu$^1$\orcid{0009-0004-4022-599X},
    Changho Kim$^{1,*}$\orcid{0000-0002-4064-8237},
    Andy J. Nonaka$^2$\orcid{0000-0003-1791-0265}, and
    Yuanran Zhu$^2$\orcid{0000-0001-6851-4161}
}

\affil{$^1$Department of Applied Mathematics, University of California, Merced, California, USA}

\affil{$^2$Lawrence Berkeley National Laboratory, Berkeley, California, USA}

\affil{$^*$Author to whom any correspondence should be addressed.}

\email{ckim103@ucmerced.edu}


\begin{abstract} 
We present a multidimensional statistics-informed neural network (SINN) framework for learning and generating stochastic trajectories of multicomponent stochastic processes. 
SINN is built upon a deterministic recurrent neural network that takes streams of white-noise sequences as input and generates ensembles of stochastic trajectories whose statistical properties match those of the underlying stochastic dynamics. 
While the capabilities of SINN have been demonstrated for various one-dimensional non-Markovian processes, extending the framework to multidimensional stochastic processes is essential for addressing real-world applications. 
In this paper, we present several key developments toward this goal. 
First, because direct estimation of multivariate probability density functions (PDFs) becomes computationally challenging with increasing dimension, we represent joint-distribution information using collections of one-dimensional PDFs constructed from linear combinations of the process components. 
Second, to accommodate the larger collection of statistical objectives required to characterize multidimensional processes, we adopt a self-adaptive loss-balancing strategy that dynamically weights the different components of the total loss function. 
Third, we introduce a reference-normalized validation framework that interprets statistical discrepancies relative to the finite-sample variability between independent target ensembles. 
We also investigate the role of the input white-noise dimension in SINN performance. 
We evaluate the proposed approach using a Langevin benchmark based on two coupled Duffing oscillators and further demonstrate its applicability to the Langevin dynamics of the Fermi--Pasta--Ulam chain and kinetic Monte Carlo simulations of a lattice Lotka--Volterra model. 
Our results demonstrate that the multidimensional SINN framework provides a practical approach for modeling multidimensional stochastic dynamics.
\end{abstract}

\section{Introduction}

Recent advances in machine learning have led to powerful data-driven approaches for modeling complex dynamical systems, with neural networks (NNs) emerging as a key tool owing to their ability to capture nonlinear dynamics and temporal dependencies.
Among these approaches, physics-informed neural networks (PINNs)~\cite{RaissiPerdikarisKarniadakis2019, KarniadakisKevrekidisLuPerdikarisWangYang2021} have achieved notable success in learning and simulating deterministic systems by approximating solutions to the governing partial differential equations (PDEs). 
By incorporating physical laws directly into the training objective, PINNs embed domain knowledge in NN models, thereby promoting consistency with the underlying physics while maintaining agreement with observed data.

Many machine-learning approaches originally developed for deterministic dynamical systems have subsequently been extended to stochastic settings. 
For example, the PINN framework has been adapted to solve inverse stochastic problems~\cite{YangZhangKarniadakis2020, ChenYangDuanJinqiaoKarniadakis2021, YangDaskalakisKarniadakis2022} and to learn stochastic differential equation (SDE) models from data~\cite{OLearyPaulsonMesbah2022}.
Similarly, the neural ordinary differential equation (ODE) framework~\cite{ChenRubanovaBettencourtDuvenaud2018} has been generalized to neural SDEs~\cite{TzenRaginsky2019, KidgerFosterLiLyons2021} and latent SDEs~\cite{LiWongChenDuvenaud2020}.
The neural-SDE framework has subsequently been extended to broader classes of stochastic systems, including jump SDEs~\cite{JiaBenson2019} and stochastic PDEs~\cite{SalviLemercierGerasimovics2022}.
In parallel, the neural-flow approach, originally developed as a solver-free alternative to neural ODEs~\cite{BilosSommerRangapuramJanuschowskiGunnemann2021}, has been extended to stochastic settings through neural stochastic flows for learning weak solutions of SDEs~\cite{KiyoharaJohnsLi2025}.
Beyond these neural-differential-equation-based approaches, data-driven methods have also been developed for learning stochastic dynamics from simulation or trajectory data. 
Examples include approaches for learning stochastic flow-map operators~\cite{ChenXiu2024}, effective SDEs with learned state-dependent drift and diffusivity~\cite{DietrichMakeevKevrekidisEvangelouBertalanReichKevrekidis2023, EvangelouCuiBelloRivasMakeevKevrekidis2024}, and generalized Langevin equations~\cite{CherchiDequidtGuillinMartzelHauretBarra2024}.

While many existing approaches are built upon differential-equation-based formulations, recurrent neural network (RNN) architectures provide an alternative data-driven framework for modeling stochastic temporal processes directly from trajectory data. 
Representative examples include stochastic recurrent networks~\cite{BayerOsendorfer2014}, variational RNNs~\cite{ChungKastnerDinhGoelCourvilleBengio2015}, and related stochastic-RNN variants~\cite{FraccaroSonderbyPaquetWinther2016, GoyalSordoniCoteKeBengio2017}, which introduce latent random variables into recurrent models to capture uncertainty and complex temporal variability.
Beyond these latent-variable recurrent models, generative adversarial networks (GANs) have also been developed for time-series generation, including the time-series GAN~\cite{YoonJarrettVanDerSchaar2019} and the causal optimal transport GAN~\cite{XuWenliangMunnAcciaio2020}.
Related normalizing-flow-based approaches have been proposed for the generative modeling of continuous stochastic processes~\cite{DengChangBrubakerMoriLehrmann2020}.
More recently, similar objectives have been pursued using transformer-based stochastic time-series models~\cite{TangMatteson2021} and diffusion-based stochastic time-series models~\cite{RasulSewardSchusterVollgraf2021, KolloviehAnsariBohlkeSchneiderZschiegnerWangWang2023}.

\begin{figure}[t]
    \centering
    \includegraphics[width=0.9\textwidth]{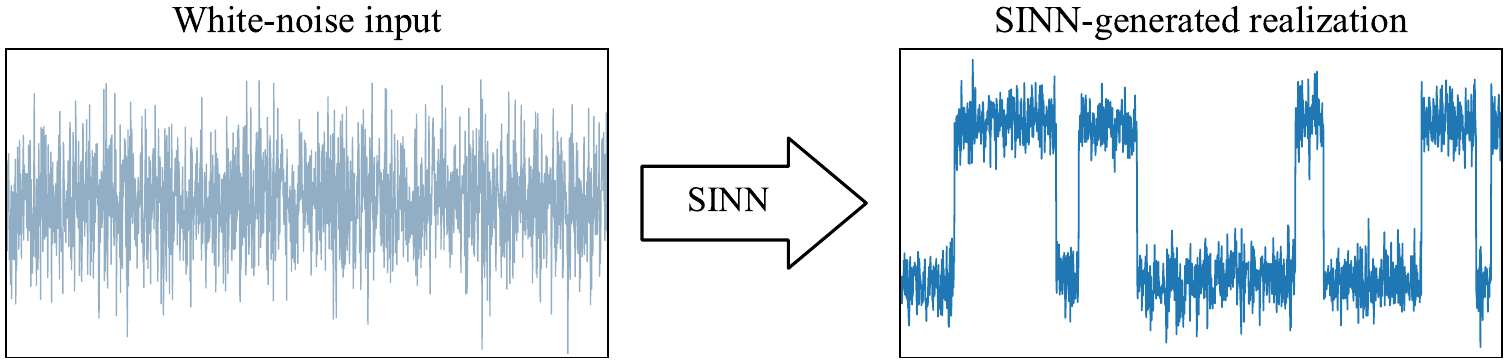}
    \caption{Basic idea of SINN. Each white-noise input sequence is transformed into a trajectory realization.
    The ensemble of generated trajectories reproduces the statistical properties of the target stochastic process.}
    \label{fig:basic_sinn}
\end{figure}

While the methods described above provide powerful frameworks for learning stochastic dynamics, many of them focus on generative modeling or on learning explicit stochastic differential equation representations.
In the context of surrogate modeling of stochastic dynamical systems, however, the objective considered here is different. 
Rather than identifying an underlying stochastic equation or modeling trajectories primarily for generative or predictive purposes, we seek to construct an efficient stochastic trajectory generator capable of producing a large number of realizations that faithfully reproduce the statistical properties of the target process. 
To address this objective, statistics-informed neural networks (SINNs) have been proposed as a data-driven framework for stochastic surrogate modeling that does not require an explicit SDE representation~\cite{ZhuTangKim2023}. 
In the SINN formulation, each generated realization of a target stochastic process is obtained by applying a deterministic recurrent-neural-network-based transformation to an input white-noise sequence (Figure~\ref{fig:basic_sinn}). 
Unlike conventional stochastic-process representations that rely on prescribed model structures or expansion bases, such as ARIMA models~\cite{BoxJenkins1970}, Karhunen--Lo\`eve expansions~\cite{PhoonHuangQuek2005}, and polynomial chaos expansions~\cite{SakamotoGhanem2002}, SINNs exploit the expressive power of recurrent neural networks to learn the transformation from data. 
This formulation is supported by universal approximation results~\cite{ZhuTangKim2023}. 
In particular, it has been shown that RNNs can approximate stochastic processes governed by stochastic differential equations through suitable transformations of random inputs.

In the original SINN formulation~\cite{ZhuTangKim2023}, long short-term memory (LSTM) units~\cite{HochreiterSchmidthuber1997} were employed to capture the complex temporal dependencies and long-range memory effects characteristic of many stochastic dynamical systems. 
The methodology was demonstrated on representative one-dimensional stochastic processes, including a coarse-grained representation of the Fermi--Pasta--Ulam chain and Brownian dynamics in a double-well potential. 
In these examples, SINNs successfully reproduced the statistical properties of the underlying processes, such as probability distributions and time autocorrelation functions. 
Notably, SINNs were shown to generate statistically faithful trajectories using time increments substantially larger than those required by conventional SDE integration schemes, highlighting their potential for coarse-grained stochastic modeling and efficient multiscale simulations. 
Since its introduction, the SINN framework and related statistics-informed ideas have been applied to a variety of stochastic modeling problems, including reduced representations of geophysical turbulence~\cite{FoldesCamporealeMarino2024}, financial volatility forecasting~\cite{LiuWang2024}, and the construction of stochastic trajectory generators for trawl processes~\cite{BelkovaMikhailov2025}.
Despite these encouraging developments, existing studies have primarily focused on one-dimensional stochastic processes, and the application of SINNs to multidimensional stochastic systems remains largely unexplored.

In this work, we develop a multidimensional SINN framework for learning and generating stochastic trajectories of multicomponent stochastic processes. 
In the SINN framework, training is performed by incorporating statistical measures that characterize the target stochastic process into the loss function, thereby encouraging the generated trajectories to reproduce the desired statistical properties. 
While this strategy is effective for one-dimensional processes, extending SINN to multidimensional stochastic systems introduces several additional challenges. 
Multidimensional stochastic dynamics are characterized not only by the statistics of individual components but also by their mutual dependencies. 
Consequently, a straightforward extension based on independently trained one-dimensional SINNs cannot reproduce the inter-component correlations and higher-order cross statistics required to describe coupled stochastic systems. 
Accounting for these dependencies requires tractable representations of joint-distribution information, since direct estimation of multivariate probability density functions (PDFs) becomes increasingly challenging with dimension.
The additional cross-component statistics also substantially enlarge the collection of statistical objectives.
Moreover, these objectives may exhibit different optimization sensitivities and convergence behaviors, making their simultaneous optimization nontrivial.
Because statistical quantities are estimated from finite trajectory ensembles, interpreting the resulting validation discrepancies also requires accounting for the finite-sample variability present even between independent target ensembles. 
Finally, the dimensionality of the driving white-noise input becomes an important modeling consideration. 
To address these issues, we develop a multidimensional SINN methodology based on projected-PDF representations of joint-distribution information, self-adaptive loss balancing, and reference-normalized validation, and we investigate the role of the white-noise input dimension. 
The proposed methodology is first assessed using a Langevin benchmark based on two coupled Duffing oscillators and then applied to the Langevin dynamics of a Fermi--Pasta--Ulam chain and a lattice Lotka--Volterra model simulated using kinetic Monte Carlo methods.

The remainder of this paper is organized as follows.
Section~\ref{sec2} presents the general SINN formulation for multidimensional stochastic processes and identifies the statistical and optimization challenges that arise in the multidimensional setting.
Section~\ref{sec3} develops the proposed methodology, including the projected-PDF representation, self-adaptive loss balancing, and the training and validation framework.
Section~\ref{sec4} evaluates the methodological components using the coupled-Duffing benchmark.
Section~\ref{sec5} applies the resulting framework to the Fermi--Pasta--Ulam chain and lattice Lotka--Volterra model, and Section~\ref{sec6} concludes the paper.

\section{Formulation of Multidimensional SINN}
\label{sec2}

In this section, we present the formulation of the SINN framework employed in this work.
To isolate the methodological issues associated with multidimensional stochastic dynamics, we consider the same class of stochastic processes as in the original SINN work~\cite{ZhuTangKim2023}: continuous-valued stochastic processes and stochastic observables that admit continuous-range approximations, with statistical properties that are invariant under shifts of the time origin and with well-defined moments up to the orders required for statistical characterization.
We use Gaussian white-noise sequences as the driving random inputs. 
Within this setting, Section~\ref{subsec:Architecture} describes the recurrent architecture used to generate multicomponent trajectories, and Section~\ref{subsec:GenLoss} formulates the statistics-informed training objective.

\subsection{Architecture}
\label{subsec:Architecture}

The SINN architecture maps white-noise input sequences to trajectory realizations of a target stochastic process.
Consider a $d_\mathrm{in}$-dimensional white-noise input sequence and its corresponding $d$-dimensional SINN-generated trajectory, both of length $T$.
We assume that the input white-noise vectors are independent and identically distributed Gaussian random vectors satisfying
\begin{equation} 
    \boldsymbol{\xi}_t \overset{\mathrm{i.i.d.}}{\sim} \mathcal{N}(\boldsymbol{0},\mathbf{I}_{d_\mathrm{in}}), \quad t=1,\ldots,T . 
\end{equation}
We denote the Gaussian white-noise input sequence and the generated trajectory by 
\begin{equation} 
    \boldsymbol{\xi}_{1:T} = \{\boldsymbol{\xi}_t\in\mathbb{R}^{d_\mathrm{in}}\}_{t=1}^T,\quad
    \widehat{\boldsymbol{x}}_{1:T} = \{\widehat{\boldsymbol{x}}_t\in\mathbb{R}^d\}_{t=1}^T,
\end{equation} 
respectively.
The network defines the sequence-to-sequence mapping
\begin{equation}
    \boldsymbol{\xi}_{1:T} \mapsto \widehat{\boldsymbol{x}}_{1:T},
\end{equation}
where the output at time $t$ depends on the current input and information from previous time steps encoded through the recurrent architecture.
This dependence can be represented schematically as
\begin{equation}
    \widehat{\boldsymbol{x}}_t=\mathrm{SINN}(\boldsymbol{\xi}_t;\boldsymbol{\xi}_{1:t-1})
    \mbox{ with } \widehat{\boldsymbol{x}}_1=\mathrm{SINN}(\boldsymbol{\xi}_1).
\end{equation}

\begin{figure*}[t]
    \centering
    \begin{minipage}[t]{0.26\textwidth}
        \raggedright
        {\small(a)}\par
        \vspace{2pt}
        \centering
        \includegraphics[width=\linewidth]{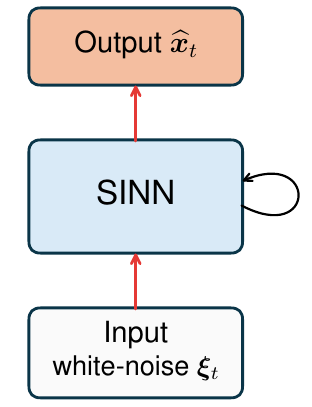}
    \end{minipage}
    \hfill
    \begin{minipage}[t]{0.70\textwidth}
        \raggedright
        {\small(b)}\par
        \vspace{2pt}
        \centering
        \includegraphics[width=\linewidth]{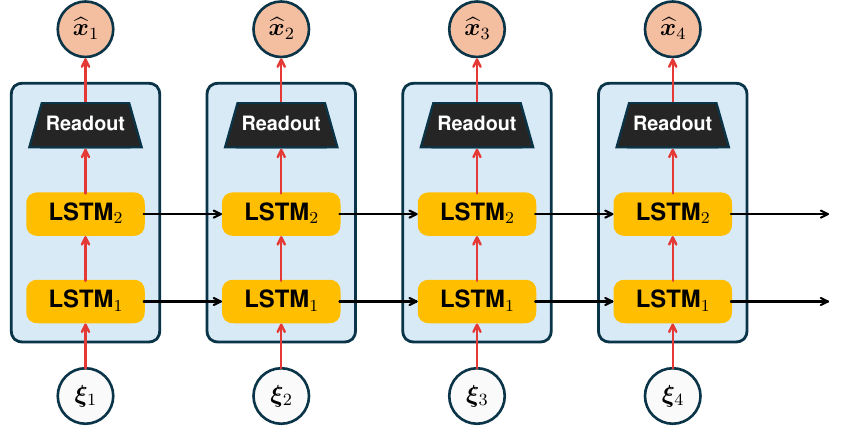}
    \end{minipage}
    \caption{Recurrent architecture of SINN. 
    (a) Time-folded representation. 
    (b) Example implementation consisting of two stacked LSTM layers and a linear readout, shown in its time-unrolled form.
    The inputs $\boldsymbol{\xi}_t \in \mathbb{R}^{d_\mathrm{in}}$ and outputs $\widehat{\boldsymbol{x}}_t \in \mathbb{R}^d$ are vector-valued. 
    The replicated blocks share the same parameters across time.
    Horizontal arrows indicate the propagation of hidden and cell states across time.}
    \label{fig:sinn_structure}
\end{figure*}

As shown by the universal approximation theorem for RNNs with Gaussian inputs~\cite{ZhuTangKim2023}, a one-layer deterministic RNN can accurately approximate the finite-difference scheme of a time-homogeneous Markovian SDE driven by Gaussian white noise, provided that the network is sufficiently wide. 
It should be noted, however, that this result concerns the representational capacity of the network and does not guarantee successful training or accurate reproduction of long-time statistical behavior in practice. 
Many stochastic processes of practical interest exhibit more complex temporal dependencies and long-range memory effects. 
To enhance the ability of SINN to represent such processes, we employ long short-term memory (LSTM) units~\cite{HochreiterSchmidthuber1997}, which are specifically designed to capture long-term temporal information. 
To further increase model capacity, multiple LSTM layers are stacked to construct the recurrent network.
Figure~\ref{fig:sinn_structure} illustrates the relationship between the compact RNN representation of SINN (panel~(a)) and an implementation using two stacked LSTM layers (panel~(b)). 
A linear readout layer is employed to map the LSTM output to the $d$-dimensional output space, since the hidden-state dimension of the LSTM does not necessarily coincide with the dimensionality of the target stochastic process.

It should be emphasized that the recurrent network itself is entirely deterministic, with stochasticity entering the model exclusively through the white-noise input sequence. 
Different input realizations therefore produce different output trajectories through the deterministic transformation defined by SINN. 
By sampling many independent white-noise sequences, the network generates an ensemble of trajectory realizations from which the statistical properties of the modeled stochastic process can be estimated.

\subsection{Loss Function and Training}
\label{subsec:GenLoss}

The overall training objective of SINN is to generate an ensemble of stochastic trajectories whose statistical properties match those of the target process. 
Training is therefore performed by minimizing discrepancies between statistical quantities of the target process and their counterparts estimated from generated trajectories. 
Unlike trajectory-wise learning approaches, which require a correspondence between individual target and generated realizations, SINN employs an ensemble-based loss function and is trained to reproduce selected statistical signatures of the stochastic process rather than individual sample paths.

To estimate the statistical quantities appearing in the loss function, an ensemble of independent white-noise input sequences is sampled and transformed by the network into an ensemble of trajectory realizations. 
We denote the white-noise ensemble, the corresponding ensemble of SINN-generated trajectories, and an ensemble of target realizations by 
\begin{equation} 
    \boldsymbol{\xi} = \{\boldsymbol{\xi}_{1:T}^{(i)}\}_{i=1}^{N},\quad 
    \widehat{\boldsymbol{x}} 
    = \widehat{\boldsymbol{x}}_{\boldsymbol{\theta}}(\boldsymbol{\xi}) 
    = \{\widehat{\boldsymbol{x}}_{1:T}^{(i)}\}_{i=1}^{N},\quad \boldsymbol{x} 
    = \{\boldsymbol{x}_{1:T}^{(i)}\}_{i=1}^{N}, 
\end{equation} 
respectively, where $N$ is the ensemble size and $\boldsymbol{\theta}$ denotes the network parameters.
Selected statistics computed from the generated ensemble $\widehat{\boldsymbol{x}}$ are then compared with the corresponding statistics precomputed from the target ensemble $\boldsymbol{x}$.

Let $\mathcal{S}$ denote the collection of statistical descriptors used to characterize the stochastic process:
\begin{equation}
    \mathcal{S} = \{s_1,s_2,\ldots,s_{|\mathcal{S}|}\}. 
\end{equation}
Each descriptor may represent either an individual statistical quantity or a collection of related quantities.
The general statistics-based objective is then written as
\begin{equation}
    \mathcal{L}_{\mathrm{total}}
    = \sum_{s\in\mathcal{S}} w_s\mathcal{L}_s,
    \label{eq:general-statistical-loss}
\end{equation}
where $w_s>0$ denotes the weight assigned to the loss associated with descriptor $s$.
The statistical loss associated with descriptor $s$ is defined as
\begin{equation}
    \mathcal{L}_s = \ell_s\!\left( s(\widehat{\boldsymbol{x}}), s(\boldsymbol{x}) \right),
\end{equation}
where $s(\widehat{\boldsymbol{x}})$ and $s(\boldsymbol{x})$ denote the evaluations of descriptor $s$ for the generated and target ensembles, respectively, and $\ell_s$ quantifies their discrepancy.

\begin{figure}[t]
    \centering
    \includegraphics[width=\textwidth]{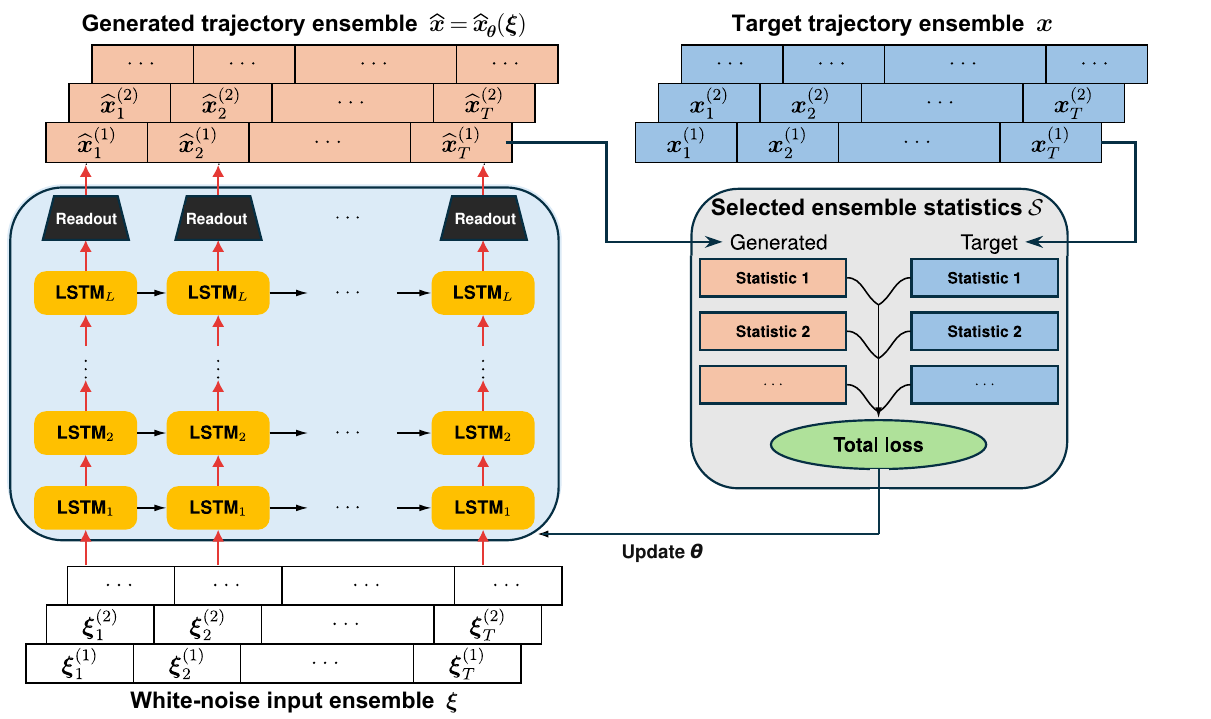}
    \caption{Training workflow of SINN with an ensemble-based loss function. 
    At each iteration, statistical quantities computed from a generated trajectory ensemble are compared with the corresponding quantities computed from the target ensemble, and the resulting loss is used to optimize the network parameters.}
    \label{fig:sinn_training}
\end{figure}

The ensemble-based formulation has two practical implications for training. 
First, because no pathwise correspondence between target and generated trajectories is required, the white-noise sequences associated with the target trajectories need not be tracked or stored; many practical data sources, including experiments and molecular-dynamics simulations, do not provide such information. 
Second, SINN uses a newly sampled batch of independent white-noise sequences at every optimization step. 
This continual resampling reduces dependence on particular input realizations and promotes generalization across white-noise samples.
Figure~\ref{fig:sinn_training} summarizes the ensemble-based training workflow described above, complementing the single-trajectory illustrations of SINN operation and architecture in Figures~\ref{fig:basic_sinn} and~\ref{fig:sinn_structure}.

The formulation above applies to both scalar and vector-valued stochastic processes. 
In the original one-dimensional SINN, the probability density function (PDF) and autocorrelation function (ACF) are used with equal loss weights, yielding $\mathcal{L}_\mathrm{total} = \mathcal{L}_\mathrm{PDF}+\mathcal{L}_\mathrm{ACF}$~\cite{ZhuTangKim2023}. 
For multidimensional target processes, the recurrent architecture and ensemble-based training procedure retain the same general form, but the statistical objective must additionally characterize dependencies among the output components. 
Componentwise PDFs and ACFs describe the marginal distributions and individual temporal behavior, but are generally insufficient to characterize cross-component dependence. 
The multidimensional objective must therefore account for both cross-component temporal correlations and joint-distribution information, while avoiding direct reliance on increasingly impractical multivariate-PDF estimation. 
Section~\ref{subsec:projected_pdf} develops the statistical representation used for this purpose.

A second challenge arises from the enlarged collection of statistical objectives required to characterize multidimensional stochastic processes. 
Because these objectives may exhibit different optimization sensitivities and convergence behaviors, their relative weights in Eq.~\eqref{eq:general-statistical-loss} can substantially influence which statistical quantities are reproduced most accurately. 
Selecting suitable fixed weights is therefore nontrivial and may become increasingly difficult as additional statistical constraints are introduced. 
Although this issue is not unique to multidimensional SINNs, it becomes particularly important when multiple componentwise and cross-component statistics must be optimized simultaneously. 
This motivates the self-adaptive loss-balancing strategy developed in Section~\ref{subsec:loss_balancing}. 
Section~\ref{subsec:optim_strat} then assembles these methodological components into the training and validation workflow used throughout this work.

\section{Multidimensional SINN Methodology}
\label{sec3}

We now develop the multidimensional SINN methodology motivated by the formulation in Section~\ref{sec2}. 
The computation of standard statistical quantities, including componentwise probability density functions (PDFs), autocorrelation functions (ACFs), and cross-correlation functions (CCFs), largely follows the original SINN formulation~\cite{ZhuTangKim2023} and standard statistical practice; their definitions and numerical implementation are provided in Appendix~\ref{app:est_stat}. 
The following subsections develop the projected-PDF representation (Section~\ref{subsec:projected_pdf}) and self-adaptive loss-balancing strategy (Section~\ref{subsec:loss_balancing}) and present the training and validation workflow (Section~\ref{subsec:optim_strat}) used in this work.

\subsection{Projected-PDF Representation of Joint Distributions}
\label{subsec:projected_pdf}

Direct incorporation of multivariate PDFs into the statistical objective becomes increasingly challenging with dimension because both the computational cost and statistical uncertainty of density estimation increase for finite trajectory ensembles. 
We therefore represent instantaneous joint-distribution information using a collection of one-dimensional PDFs constructed from linear combinations of the process components, while temporal dependence is characterized separately through the ACFs and CCFs included in the statistical objective.

The mathematical motivation for this approach is provided by the Cram\'er--Wold theorem~\cite{CramerWold1936}, which states that a multivariate probability distribution is uniquely determined by the distributions of all of its one-dimensional linear projections. 
For example, consider two random vectors $\boldsymbol{X}=(X_1,X_2)\in\mathbb{R}^2$ and $\widehat{\boldsymbol{X}}=(\widehat{X}_1,\widehat{X}_2)\in\mathbb{R}^2$. 
According to the theorem, $\boldsymbol{X}$ and $\widehat{\boldsymbol{X}}$ have the same distribution if and only if $\boldsymbol{w}\cdot\boldsymbol{X} = w_1 X_1 + w_2 X_2$ and $\boldsymbol{w}\cdot\widehat{\boldsymbol{X}} = w_1\widehat{X}_1 + w_2\widehat{X}_2$ have the same distribution for every $\boldsymbol{w}=(w_1,w_2)\in\mathbb{R}^2$. 
In practice, however, we employ only a finite set of projection directions. 
Consequently, the projected-PDF representation provides only partial information about the joint distribution, while remaining computationally tractable.

Each projected variable corresponds to a one-dimensional projection of the multidimensional process along a selected direction. 
For a selected unit vector $\boldsymbol{w}\in\mathbb{R}^d$, we define the projected variables associated with the target and generated processes as
\begin{equation} 
    u_{\boldsymbol{w},t}
    = \frac{\boldsymbol{w}\cdot\boldsymbol{x}_t}{\varsigma_{\boldsymbol{w}}}, \quad
    \widehat{u}_{\boldsymbol{w},t} 
    = \frac{\boldsymbol{w}\cdot\widehat{\boldsymbol{x}}_t}{\varsigma_{\boldsymbol{w}}},
\label{eq:projected_variable} 
\end{equation} 
Here, the normalization factor $\varsigma_{\boldsymbol{w}}$ is the standard deviation of the projected target process in the direction $\boldsymbol{w}$, estimated from the target data, and is used for both the target and generated projected variables, thereby placing different projections on comparable scales.

The PDFs of the projected processes $\{u_{\boldsymbol{w},t}\}_{t=1}^T$ and $\{\widehat{u}_{\boldsymbol{w},t}\}_{t=1}^T$ and their corresponding losses are evaluated using the same one-dimensional procedure described in Appendix~\ref{app:est_stat}. 
Increasing the number of projection directions can capture more information about the underlying joint distribution and provide a more stringent comparison between the target and generated processes, but at increased computational cost. 
The number and choice of projection directions may therefore depend on the target process and application. 
In the following, we develop a general approach for selecting a moderate collection of informative projection directions for multidimensional SINN training.

Although the projected-PDF objective compares probability densities directly, agreement between projected PDFs also constrains their moments and cumulants, provided that the required moments exist.
We use cumulants as an analytical tool to characterize the cross-component information encoded by the projected PDFs and to guide the selection of projection directions. 
This idea is first illustrated using the second-order cumulant, namely the variance. 
For $\boldsymbol{X}\in\mathbb{R}^d$ and a projection direction $\boldsymbol{w}\in\mathbb{R}^d$, the variance of the unnormalized projection satisfies 
\begin{equation} 
    \operatorname{Var}\!\left(\boldsymbol{w}\cdot\boldsymbol{X}\right) 
    = \sum_i w_i^2\operatorname{Var}(X_i) + 2\sum_{i<j}w_iw_j\operatorname{Cov}(X_i,X_j).
\end{equation} 
Thus, a projection direction with multiple nonzero components contains information about cross-covariances that is not available from the componentwise PDFs alone. 
More generally, cumulants of projected variables can be expressed as linear combinations of the corresponding joint cumulants of $\boldsymbol{X}$~\cite{McCullagh1987}: 
\begin{equation} 
    \kappa_m\!\left(\boldsymbol{w}\cdot\boldsymbol{X}\right) 
    = \sum_{|\alpha|=m} \frac{m!}{\alpha!}\, \boldsymbol{w}^{\alpha}\, \kappa_{\alpha}(\boldsymbol{X}),
\label{eq:cumulant_multilinearity}
\end{equation} 
where $\alpha$ is a multi-index satisfying $|\alpha|=m$ and $\kappa_{\alpha}(\boldsymbol{X})$ is the corresponding joint cumulant. 
Equation~\eqref{eq:cumulant_multilinearity} shows that each projection direction constrains a different linear combination of the order-$m$ joint cumulants, making the selection of projection directions an important practical consideration. 
For the following cumulant analysis, we consider the unnormalized projections $\boldsymbol{w}\cdot\boldsymbol{X}$. 
The normalization in Eq.~\eqref{eq:projected_variable} introduces only direction-dependent scaling and therefore does not affect the rank properties discussed below.

To balance information content and computational cost, we construct the projection directions using collections of orthonormal frames. 
Each frame consists of $d$ mutually orthogonal unit vectors in $\mathbb{R}^d$ and can therefore be represented by an orthogonal matrix $\mathbf{W}$ whose rows are the projection directions $\boldsymbol{w}$. 
If $n_\mathrm{f}$ denotes the number of frames, including the coordinate frame, the complete collection used in the cumulant analysis contains $N_\mathrm{dir}=d\,n_\mathrm{f}$ directions. 
Because the coordinate directions are already represented by the componentwise PDFs, only the $d(n_\mathrm{f}-1)$ directions from the remaining frames introduce additional projected-PDF loss terms.
The specific frames used in this work are chosen to provide complementary directions while maintaining favorable rank and conditioning properties. 
Details of the associated rank and conditioning analyses are provided in Appendix~\ref{app:projection_directions}.

For $d=2$, we employ $n_\mathrm{f}=4$ frames. 
The first frame is $\mathbf{W}^{(1)}=\mathbf{I}_2$, and the remaining three frames are 
\begin{equation} 
    \mathbf{W}^{(2)} = 
    \begin{bmatrix} 
        \frac{1}{\sqrt{2}} & \frac{1}{\sqrt{2}} \\ 
        \frac{1}{\sqrt{2}} & -\frac{1}{\sqrt{2}} 
    \end{bmatrix}, \quad 
    \mathbf{W}^{(3)} = 
    \begin{bmatrix} 
        \frac{1}{\sqrt{5}} & \frac{2}{\sqrt{5}} \\ 
        \frac{2}{\sqrt{5}} & -\frac{1}{\sqrt{5}} 
    \end{bmatrix}, \quad 
    \mathbf{W}^{(4)} = 
    \begin{bmatrix} 
        \frac{1}{\sqrt{5}} & -\frac{2}{\sqrt{5}} \\ 
        \frac{2}{\sqrt{5}} & \frac{1}{\sqrt{5}} 
    \end{bmatrix}. 
    \label{eq:frames2d} 
\end{equation}
The coordinate frame $\mathbf{W}^{(1)}$ is included in the cumulant analysis but introduces no additional projected-PDF terms; the remaining three frames contribute six such terms. 
As shown in Appendix~\ref{app:projection_directions}, the complete set of eight directions yields full-column-rank cumulant coefficient matrices through order seven, so all two-dimensional joint cumulants through that order are uniquely identifiable from the selected projections.

For $d=3$, we employ $n_\mathrm{f}=4$ frames. 
The first frame is $\mathbf{W}^{(1)}=\mathbf{I}_3$, and the remaining three frames are 
\begin{equation} 
    \mathbf{W}^{(2)} = 
    \begin{bmatrix} 
        \frac{1}{\sqrt{3}} & \frac{1}{\sqrt{3}} & \frac{1}{\sqrt{3}} \\ 
        \frac{1}{\sqrt{2}} & -\frac{1}{\sqrt{2}} & 0 \\ 
        \frac{1}{\sqrt{6}} & \frac{1}{\sqrt{6}} & -\frac{2}{\sqrt{6}} 
    \end{bmatrix}, \, 
    \mathbf{W}^{(3)} = 
    \begin{bmatrix} 
        \frac{1}{\sqrt{3}} & \frac{1}{\sqrt{3}} & -\frac{1}{\sqrt{3}} \\ 
        0 & \frac{1}{\sqrt{2}} & \frac{1}{\sqrt{2}} \\ 
        \frac{2}{\sqrt{6}} & -\frac{1}{\sqrt{6}} & \frac{1}{\sqrt{6}} 
    \end{bmatrix}, \, 
    \mathbf{W}^{(4)} = 
    \begin{bmatrix} 
        \frac{1}{\sqrt{3}} & -\frac{1}{\sqrt{3}} & \frac{1}{\sqrt{3}} \\ 
        \frac{1}{\sqrt{2}} & \frac{1}{\sqrt{2}} & 0 \\ 
        \frac{1}{\sqrt{6}} & -\frac{1}{\sqrt{6}} & -\frac{2}{\sqrt{6}} 
    \end{bmatrix}. 
    \label{eq:frames3d} 
\end{equation} 
The coordinate frame $\mathbf{W}^{(1)}$ is included in the cumulant analysis but introduces no additional projected-PDF terms; the remaining three frames contribute nine such terms. 
As shown in Appendix~\ref{app:projection_directions}, the complete set of twelve directions yields full-column-rank cumulant coefficient matrices through order three, so all three-dimensional joint cumulants through that order are uniquely identifiable from the selected projections. 
The resulting two- and three-dimensional direction sets are used throughout this work to construct the projected-PDF objectives.

\subsection{Self-Adaptive Loss Balancing}
\label{subsec:loss_balancing}

The multiple statistical objectives entering the multidimensional SINN loss can exhibit different optimization sensitivities and convergence behaviors, making suitable fixed weights difficult to determine. 
We therefore adopt the self-adaptive loss-balancing framework of Ref.~\cite{xiang2022self}, originally developed for physics-informed neural networks (PINNs). 
In PINNs, data-mismatch, boundary-condition, and PDE-residual losses may contribute unequally to the optimization process; an analogous difficulty arises in multidimensional SINNs, where multiple statistical objectives must be satisfied simultaneously. 
The method introduces trainable scale parameters that are optimized jointly with the network parameters, allowing the relative contributions of the statistical loss components to evolve automatically during training.

Specifically, consider a collection of statistical loss components $\{\mathcal{L}_s(\boldsymbol{\theta})\}_{s\in\mathcal{S}}$. 
The loss-balancing framework is agnostic to how the statistical quantities are organized: each component may represent either an individual statistical objective or an aggregate of several related objectives. 
The particular organization adopted in this work is described in Section~\ref{subsec:optim_strat}. 
A trainable scale parameter $\sigma_s$ is introduced for each loss component $s\in\mathcal{S}$, yielding the loss-balanced objective 
\begin{equation} 
    \mathcal{L}_\mathrm{LB}(\boldsymbol{\theta},\boldsymbol{\sigma}) 
    = \sum_{s\in\mathcal{S}} \left[ \frac{\mathcal{L}_s(\boldsymbol{\theta})}{2\sigma_s^2}
    + \log\sigma_s \right], \quad 
    \boldsymbol{\sigma} =\{\sigma_s\}_{s\in\mathcal{S}}. 
\label{eq:lb_loss_fn} 
\end{equation} 
Equation~\eqref{eq:lb_loss_fn} may be viewed as a Gaussian-likelihood-motivated objective in which each statistical loss component is associated with a trainable positive scale parameter $\sigma_s$. 
The factor $1/(2\sigma_s^2)$ acts as an adaptive weight for $\mathcal{L}_s$, while the logarithmic term prevents the scales from increasing without bound and thereby excludes the trivial solution in which all effective loss weights vanish. 
To ensure positivity, we introduce unconstrained trainable variables $\boldsymbol{\alpha}=\{\alpha_s\}_{s\in\mathcal{S}}$ and parameterize the scales as $\sigma_s=\exp(\alpha_s)$.

During training, the network parameters $\boldsymbol{\theta}$ and the scale coordinates $\boldsymbol{\alpha}$ are optimized simultaneously. 
At each optimization step, a fresh batch of white-noise sequences is transformed into an ensemble of generated trajectories, from which the statistical loss components $\{\mathcal{L}_s\}_{s\in\mathcal{S}}$ and the loss-balanced objective $\mathcal{L}_\mathrm{LB}$ are evaluated. 
The resulting gradients are used to update both $\boldsymbol{\theta}$ and $\boldsymbol{\alpha}$, with the adaptive weights determined by the corresponding scale parameters $\sigma_s=\exp(\alpha_s)$.

\subsection{Optimization Strategy}
\label{subsec:optim_strat}

We now assemble the methodological components introduced above into the training and validation workflow used throughout this work.

\paragraph{Warm-up Strategy.}

Because SINN employs a recurrent architecture, the initial portion of a generated trajectory may be influenced by the hidden-state initialization rather than the stationary dynamics of interest.
Accordingly, sequences of length $2T$ are generated, the first $T$ outputs are discarded as a warm-up period, and the remaining $T$ time steps are retained for statistical analysis.
The warm-up period is chosen sufficiently long so that its influence on the statistics computed from the retained trajectory segment is negligible. 
Throughout this paper, the retained portion is identified with the trajectory $\widehat{\boldsymbol{x}}_{1:T}$ introduced in Section~\ref{subsec:Architecture}; that is, the indexing convention is shifted after discarding the warm-up segment.

\paragraph{Statistical Objective.}

For the multidimensional SINN models considered in this work, related statistical quantities are grouped into family-level losses rather than assigned separate adaptive weights individually. 
Each family is associated with a single aggregate loss and a single adaptive weight, reducing the number of adaptive weighting parameters while retaining distinct objectives for different types of statistical information. 
Specifically, we employ the family collection 
\begin{equation} 
    \mathcal{S} = \{ \mathrm{PDF}, \mathrm{PROJ}, \mathrm{ACF}, \mathrm{CCF}, \mathrm{ACF2} \},
\label{eq:stat_family} 
\end{equation} 
where $\mathrm{PDF}$ contains $d$ marginal PDFs, $\mathrm{PROJ}$ contains $N_\mathrm{proj}$ projected PDFs, $\mathrm{ACF}$ contains $d$ time correlation functions, $\mathrm{CCF}$ contains $d(d-1)$ time cross-correlation functions, and $\mathrm{ACF2}$ contains $d$ time correlation functions of squared process components. 
The $\mathrm{ACF2}$ family is included to capture higher-order temporal structure not represented by the ordinary ACFs. 
The numerical procedures used to estimate these quantities are described in Appendix~\ref{app:est_stat}.

\paragraph{Loss Evaluation.} 

For each statistical family, the discrepancy function $\ell_s$ may take either an absolute or relative form, depending on the numerical study. 
For $\|s(\boldsymbol{x})\|_2>0$, the relative loss is defined as 
\begin{equation} 
    \mathcal{L}_s^{\mathrm{rel}} 
    = \frac{\bigl\|s(\widehat{\boldsymbol{x}})-s(\boldsymbol{x})\bigr\|_2^2} 
    {\bigl\|s(\boldsymbol{x})\bigr\|_2^2}. 
\label{eq:loss_term} 
\end{equation} 
This normalization yields a dimensionless discrepancy by accounting for the overall magnitude of the target statistic, whereas the corresponding absolute loss uses the mean squared error (MSE) discrepancy without this normalization. 
The loss form used in each numerical study is specified in its experimental setup, and the family-specific definitions are provided in Appendix~\ref{app:est_stat}. 
In the loss-balanced objective of Eq.~\eqref{eq:lb_loss_fn}, $\mathcal{L}_s$ denotes the loss form selected for family $s$.

\paragraph{Optimization Procedure.}

The overall multidimensional SINN training workflow combines the warm-up strategy, family-based statistical objective, and self-adaptive loss-balancing framework described above.
Algorithm~\ref{alg:lb} summarizes the resulting optimization procedure.

\begin{algorithm}[htbp]
\caption{Self-adaptive loss-balanced training procedure for SINN}
\label{alg:lb}
\begin{algorithmic}[1]
    \Require Target ensemble $\boldsymbol{x}$; statistical families $\mathcal{S}$; family losses $\{\mathcal{L}_s\}_{s\in\mathcal{S}}$; iteration budget \textsc{MaxIters}; learning rates $\eta_{\boldsymbol{\theta}}$ and $\eta_{\boldsymbol{\alpha}}$
    \State Precompute the target statistics required by the family losses
    \State Initialize $\boldsymbol{\theta}_0$ and the scale coordinates $\boldsymbol{\alpha}_0$, with $\boldsymbol{\sigma}_0 = \boldsymbol{\sigma}(\boldsymbol{\alpha}_0)>0$
    \For{$i=0,\ldots,\textsc{MaxIters}-1$}
        \State Draw a fresh batch of white-noise sequences of length $2T$
        \State Pass them through the SINN and retain the final $T$ output steps as $\widehat{\boldsymbol{x}}_i$
        \State Compute $\mathcal{L}_{s,i}$ from $\widehat{\boldsymbol{x}}_i$ and $\boldsymbol{x}$ for all $s\in\mathcal{S}$
        \State Form $\mathcal{L}_{\mathrm{LB},i}$ from Eq.~\eqref{eq:lb_loss_fn} using $\{\mathcal{L}_{s,i}\}_{s\in\mathcal{S}}$ and $\boldsymbol{\sigma}_i$
        \State Compute the gradients with respect to $\boldsymbol{\theta}$ and $\boldsymbol{\alpha}$ at
         $(\boldsymbol{\theta}_i,\boldsymbol{\alpha}_i)$
        \State Update to $(\boldsymbol{\theta}_{i+1},\boldsymbol{\alpha}_{i+1})$ with the respective Adam optimizers~\cite{KingmaBa2015}
        \State Set $\boldsymbol{\sigma}_{i+1} = \boldsymbol{\sigma}(\boldsymbol{\alpha}_{i+1})>0$
    \EndFor
\end{algorithmic}
\end{algorithm}

\paragraph{Reference Discrepancy and Validation Baseline.} 

An important issue in interpreting statistical losses is that their absolute magnitudes do not directly indicate whether a SINN-generated ensemble is statistically close to the target process.
Because all statistical quantities are estimated from finite ensembles, even two independent ensembles drawn from the same underlying distribution generally produce nonzero discrepancies. 
Consequently, a validation loss cannot be interpreted in isolation.

To separate optimization, validation, finite-sample benchmarking, and final-evaluation roles, we introduce four target ensembles. 
The fitting ensemble $\boldsymbol{x}^\mathcal{F}\equiv\boldsymbol{x}$ is used to construct the target statistics appearing in the optimization objective, while an independent validation ensemble $\boldsymbol{x}^\mathcal{V}$ is used for model selection and hyperparameter tuning. 
A second independent ensemble $\boldsymbol{x}^\mathcal{R}$ is reserved for estimating finite-sample variability and is chosen to have the same size as the generated ensemble $\widehat{\boldsymbol{x}}$.
The validation ensemble $\boldsymbol{x}^\mathcal{V}$ may have the same size as, or be larger than, $\widehat{\boldsymbol{x}}$. 
Finally, when available, a much larger ensemble $\boldsymbol{x}^\mathcal{T}$ may be used to provide highly resolved benchmark statistics or independent final-evaluation statistics.

For a statistical family $s\in\mathcal{S}$, we define the validation loss as 
\begin{equation} 
    \mathcal{L}_s^\mathrm{val} = 
    \ell_s \!\left( s(\widehat{\boldsymbol{x}}), s(\boldsymbol{x}^\mathcal{V}) \right), 
\end{equation} 
and the reference discrepancy as 
\begin{equation} 
    \mathcal{L}_s^\mathrm{ref} 
    = \ell_s \!\left( s(\boldsymbol{x}^\mathcal{R}), s(\boldsymbol{x}^{\mathcal V}) \right).
\end{equation}
The reference discrepancy measures the difference between two independent target ensembles and therefore provides a baseline for the finite-sample variability associated with the statistical estimator and the ensemble sizes involved. 
Because $\boldsymbol{x}^\mathcal{R}$ and $\widehat{\boldsymbol{x}}$ have the same ensemble size, $\mathcal{L}_s^\mathrm{ref}$ and $\mathcal{L}_s^\mathrm{val}$ compare equal-sized ensembles against the same validation target.

The ratio $\mathcal{L}_s^\mathrm{val}/\mathcal{L}_s^\mathrm{ref}$ therefore provides a dimensionless measure of performance relative to the variability expected between independent target ensembles. 
Values of order unity indicate that the discrepancy between the generated and validation ensembles is comparable to the discrepancy naturally arising between two independent finite ensembles drawn from the target process. 
Further reductions in the validation loss may still occur, but should be interpreted in the context of the finite-sample variability represented by the reference discrepancy.

For benchmark problems where a large target ensemble $\boldsymbol{x}^\mathcal{T}$ is available, additional comparisons may be performed against the highly resolved statistics estimated from $\boldsymbol{x}^{\mathcal{T}}$. 
Nevertheless, the reference discrepancy remains useful because it provides an ensemble-size-dependent statistical baseline even when highly resolved reference statistics are available.

\section{Numerical Validation}
\label{sec4}

This section validates the proposed multidimensional SINN methodology using a controlled benchmark based on two coupled Duffing oscillators. 
The benchmark is particularly suitable for assessing multidimensional SINNs because the coupling strength can be varied systematically, allowing controlled adjustment of the dependence between process components, while the stationary configurational distribution is available analytically up to a normalization constant.
The latter property enables accurate distributional reference statistics to be constructed independently of trajectory sampling.

Section~\ref{subsec:4benchmark} presents the benchmark problem, and Section~\ref{subsec:4setup} describes the experimental setup. 
We then investigate the effects of self-adaptive loss balancing (Section~\ref{subsec:4lb}), projected-PDF constraints (Section~\ref{subsec:4proj}), and the white-noise input dimension (Section~\ref{subsec:4din}).

\subsection{Coupled Duffing Benchmark}
\label{subsec:4benchmark}

As a benchmark problem, we consider the Langevin dynamics of two coupled unit-mass Duffing oscillators~\cite{KovacicBrennan2011}.
Let $q_i$ and $p_i$ denote the position and momentum of oscillator $i$ ($i=1,2$), respectively.
The dynamics are governed by 
\begin{equation}
    \dot q_i = p_i, \quad
    \dot p_i = -\frac{\partial V}{\partial q_i} -\gamma_i p_i + \sqrt{\frac{2\gamma_i}{\beta}}\,\xi_i(t), \quad 
    i=1,2.
\label{eq:dimer}
\end{equation}
Here, $\xi_1$ and $\xi_2$ are independent standard Gaussian white noises, $\gamma_i$ is the damping coefficient, and $\beta$ is the inverse temperature.
The system potential is given by
\begin{equation}
    V(q_1,q_2) =
    \left(\frac{\alpha}{2}q_1^2+\frac{\kappa_1}{4}q_1^4\right)
    +\left(\frac{\alpha}{2}q_2^2+\frac{\kappa_2}{4}q_2^4\right)
    +\frac{\lambda}{2}\bigl(q_1-q_2\bigr)^2,
\label{eq:dimerpot}
\end{equation}
where $\alpha$ specifies the harmonic contribution, $\kappa_1$ and $\kappa_2$ determine the quartic anharmonicity of the individual oscillators, and $\lambda$ is the spring constant coupling the two oscillators.
With the fluctuation--dissipation scaling in Eq.~\eqref{eq:dimer}, the invariant phase-space density is proportional to $\exp[-\beta\{V(q_1,q_2)+(p_1^2+p_2^2)/2\}]$~\cite{LeimkuhlerMatthews2013}. 
Integrating out the momenta yields the stationary configurational density
\begin{equation} 
    \pi(q_1,q_2) = \frac{\exp\bigl[-\beta V(q_1,q_2)\bigr]}{Z_q}, 
    \label{eq:gibbs} 
\end{equation} 
where $Z_q$ is the normalization constant.
Hence, Eq.~\eqref{eq:gibbs} enables distributional statistics to be evaluated independently of trajectory sampling through numerical quadrature.

In the present benchmark, SINN is trained to model the two-dimensional position process $\boldsymbol{X}_t=(q_1(t),q_2(t))$.
The benchmark is designed to examine how multidimensional SINN performs under different levels of cross-component dependence while avoiding unnecessary symmetries in the target process.
The parameter values used throughout this section are summarized in Table~\ref{tab:4params}.
Two considerations play a particularly important role in the benchmark design.

To avoid identical projected densities arising from exchange symmetry, we use unequal quartic coefficients, $\kappa_1=1$ and $\kappa_2=0.5$. 
With equal quartic coefficients, exchanging the two oscillator coordinates would leave the configurational density unchanged, causing certain projection directions to yield identical distributions. 
We also choose unequal damping coefficients, $\gamma_1 = 1$ and $\gamma_2 = 0.5$, so that the two components exhibit different relaxation behavior while preserving the same stationary configurational distribution.

A second consideration is the coupling parameter $\lambda$. 
By controlling the interaction strength between the oscillators, $\lambda$ provides a systematic way to adjust the degree of statistical dependence between the two components of $\boldsymbol{X}_t$. 
Throughout this validation study, we consider $\lambda=0$, 0.5, 1, 2, 4, and 8.

For detailed comparisons, we focus on $\lambda=4$, which exhibits the largest departure from a Gaussian-copula reference~\cite{Nelsen2006} having the same component marginals and Pearson correlation. 
This choice highlights distributional dependence beyond that captured by these quantities alone. 
The diagnostic used to quantify this departure is described in Section~S1.1 of the Supplementary Material.

\begin{table}[tbp]
\centering
    \caption{Parameters of the coupled Duffing benchmark.}
    \label{tab:4params}
    \begin{tabular}{ll}
    \hline
    Parameter & Value \\
    \hline
    Harmonic coefficient         & $\alpha=1$ \\
    Quartic coefficients         & $\kappa_1 = 1$, $\kappa_2=0.5$ \\
    Inverse temperature          & $\beta=1$ \\
    Damping coefficients         & $\gamma_1 = 1$, $\gamma_2=0.5$ \\
    Coupling parameter           & $\lambda\in\{0,0.5,1,2,4,8\}$ \\
    Integrator                   & BAOAB, $dt=10^{-2}$ \\
    Sampling interval            & $\Delta t=0.1$ \\
    Configurational quadrature grid & $[-8,8]^2$, $2001\times2001$ points \\
    Analysis window              & $T=200$ \\
    Trajectories per $\lambda$   & $N_\mathrm{tot}=20{,}000$ \\
    Fitting ensemble             & $N=400$ \\
    Retained correlation lags    & $K=200$ ($k=0,\ldots,199$) \\
    Standardized KDE grid        & $[-5,5]$, $N_z=513$ \\
    One-dimensional bandwidth    & $h=(TN)^{-1/5}=0.10456$ \\
    Two-dimensional bandwidth    & $h_2=(TN)^{-1/6}=0.15234$ \\
    Independent runs per configuration      & $20$ at $\lambda=4$, $10$ otherwise \\
    \hline
\end{tabular}
\end{table}

\subsection{Experimental Setup}
\label{subsec:4setup}

\subsubsection{Data Generation and Partition}
\label{subsubsec:4data}

Target trajectories are generated by integrating Eq.~\eqref{eq:dimer} with the BAOAB splitting scheme~\cite{LeimkuhlerMatthews2013} using a time step of $dt=10^{-2}$. 
Every tenth state is retained, yielding a sampling interval of $\Delta t=0.1$.
Initial conditions are sampled from a numerical approximation of the invariant phase-space distribution: positions follow the quadrature representation of Eq.~\eqref{eq:gibbs}, while momenta are drawn independently from the corresponding Maxwell distribution.

For each coupling strength $\lambda$, we generate $N_\mathrm{tot}=20{,}000$ independent trajectories, each containing $T=200$ retained states. 
To enable paired comparisons across coupling strengths and reduce the variability associated with Monte Carlo sampling, trajectories with the same index use common random numbers for both the initial conditions and the stochastic forcing.

Following the ensemble roles defined in Section~\ref{subsec:optim_strat}, each target ensemble is partitioned by trajectory index into a fitting ensemble $\mathcal F$, nine reference blocks $\mathcal R_1,\ldots,\mathcal R_9$, a validation ensemble $\mathcal V$, and a final-evaluation ensemble $\mathcal T$. 
The fitting ensemble and each reference block contain $N=400$ trajectories, while $|\mathcal V|=4{,}000$ and $|\mathcal T|=12{,}000$. 
All subsets are disjoint.
Preliminary experiments indicated that $N=400$ provides a reasonable balance between statistical-estimation accuracy and computational cost.
The nine reference blocks provide replicate estimates of the reference discrepancy $\mathcal{L}_s^\mathrm{ref}$ introduced in Section~\ref{subsec:optim_strat}.
Using several reference ensembles reduces sensitivity to sampling variability and yields a more robust estimate of $\mathcal{L}_s^\mathrm{ref}$.

\subsubsection{Training Configuration}
\label{subsubsec:4train}

Unless otherwise stated, all runs use two stacked LSTM layers with ten hidden units per layer and a linear readout. 
Each network is trained for $10^4$ iterations using the Adam optimizer~\cite{KingmaBa2015} with learning rate $10^{-3}$, $\beta_1=0.9$, and $\beta_2=0.999$. 
At each iteration, a fresh ensemble of $N=400$ white-noise input sequences is generated, and the retained analysis windows are extracted according to the warm-up procedure described in Section~\ref{subsec:optim_strat}.

All statistical families use the relative loss defined in Section~\ref{subsec:optim_strat}, except that at $\lambda=0$ the CCF family uses its absolute loss because the population cross-correlation vanishes identically. 
When self-adaptive loss balancing is enabled, all $\alpha_s$ are initialized to zero, which yields $\sigma_s=1$ at the first iteration. 
The $\alpha_s$ are updated using a separate Adam optimizer with the same learning rate as the network optimizer. 
When loss balancing is disabled, the applicable family losses are summed with equal weights.

\subsubsection{Model Selection and Evaluation}
\label{subsubsec:4eval}

\paragraph{Validation and checkpoint selection.}

Validation is performed every $200$ iterations using a newly generated ensemble of $N=400$ SINN trajectories obtained from a white-noise stream independent of that used for training. 
The ACF, CCF, and ACF2 losses are evaluated against statistics estimated from $\mathcal V$. 
For the PDF and PROJ families, we replace the empirical validation densities in the general definition of Section~\ref{subsec:optim_strat} with target densities computed from Eq.~\eqref{eq:gibbs} by numerical quadrature, transformed to standardized coordinates, and convolved with the same Gaussian kernel used in the corresponding KDE estimators.

The checkpoint minimizing the unweighted sum of the applicable $\mathcal{L}_s^\mathrm{val}$ values is retained. 
For consistent comparison across $n_\mathrm{f}$, the PROJ validation loss uses all six non-coordinate directions in Eq.~\eqref{eq:frames2d}, regardless of which directions are included in the training objective.
This choice evaluates all models against a common projected-distribution benchmark, regardless of the projected-PDF constraints used during training.
At $\lambda=0$, the two Duffing oscillators are uncoupled and independent, so the cross-correlation function vanishes identically. 
The CCF family remains in the training objective through its absolute loss but is excluded from the validation score and the validation ratios.

For each family included in the validation comparison, $\mathcal{L}_s^\mathrm{ref}$ is defined as the median of the nine losses obtained by replacing the generated ensemble with each reference block $\mathcal{R}_i$ while retaining the same validation target.
We define 
\begin{equation}
    r_s^\mathrm{val} =
    \frac{\mathcal{L}_s^\mathrm{val}}{\mathcal{L}_s^\mathrm{ref}}.
\label{eq:family_validation_ratio}
\end{equation}
The ratio compares the validation discrepancy with the finite-sample discrepancy obtained from an independent target ensemble of the same size as the generated ensemble, using the same validation target and loss-evaluation procedure.
Values substantially larger than unity therefore indicate that the SINN validation discrepancy exceeds the finite-sample variability represented by the reference ensembles.
We define the overall validation diagnostic as the largest applicable family-level ratio, 
\begin{equation} 
    r^\mathrm{val} = \max_s r_s^\mathrm{val}, 
\label{eq:overall_validation_ratio} 
\end{equation} 
where the maximum is taken over four families at $\lambda=0$ and all five families otherwise.
In this work, a run is classified as validation-unsuccessful, or equivalently as a validation failure, if $r^{\mathrm{val}}>10$ at its retained checkpoint, indicating that at least one family-level validation loss exceeds its reference level by more than one order of magnitude. 
Runs with $r^{\mathrm{val}}\leq10$ are referred to as validation-successful, although this classification indicates only that the failure criterion is not met. 
Values of $r^{\mathrm{val}}>10$ were not observed among the reference ensembles themselves, supporting the use of this threshold as a practical indicator of substantial disagreement with the target process.

\paragraph{Final evaluation for the input-dimension study.}

For the input-dimension study in Section~\ref{subsec:4din}, each retained model is evaluated using a newly generated independent ensemble of $N=400$ SINN trajectories. 
The final-evaluation ensemble $\mathcal{T}$ serves as the target ensemble for the correlation-based families.
As in validation, the PDF and PROJ families use target densities computed from Eq.~\eqref{eq:gibbs} by numerical quadrature and processed using the same standardization and smoothing procedures.
The componentwise standardization constants, projection scales, density grids, and bandwidths remain fixed throughout the evaluation procedure. 
We report the median across runs of the unweighted total evaluation loss, 
\begin{equation} 
    \mathcal{L}_\mathrm{total}^\mathrm{eval} = \sum_s \mathcal{L}_s^\mathrm{eval}, 
\label{eq:total_evaluation_loss} 
\end{equation} 
where the sum includes the same families as the validation score.

\subsubsection{Implementation Details}

All experiments are implemented in PyTorch~\cite{PaszkeEtAl2019} and executed on the UC Merced Pinnacles cluster. 
Each compute node includes two Intel Xeon Gold 6330 processors, $256$~GB of RAM, and NVIDIA A100 80~GB PCIe GPUs. 
Experiments use PyTorch 2.0.1 with CUDA 11.8.
Reported per-iteration timings are measured on a single NVIDIA A100 80~GB PCIe GPU with the corresponding compute node reserved exclusively.

\subsection{Self-Adaptive Loss Balancing}
\label{subsec:4lb}

We assess the performance of the self-adaptive loss-balancing strategy by comparing otherwise identical SINN models trained with and without loss balancing. 
Throughout this subsection, we fix the white-noise input dimension at $d_\mathrm{in}=2$ and use $n_\mathrm{f}=4$ projection frames in total.
The effects of these two modeling choices are examined separately in Sections~\ref{subsec:4proj} and~\ref{subsec:4din}.
For each of the balancing-enabled and balancing-disabled configurations, we perform 10 independent runs at $\lambda=0$, $0.5$, $1$, $2$, and $8$, and 20 runs at $\lambda=4$, yielding 70 runs per configuration. 
We use the larger number of runs at $\lambda=4$ for the detailed analysis below.

We begin with an overall comparison of the unweighted training losses near the end of training and the validation-successful run counts across the six coupling strengths. 
With loss balancing, the unweighted training losses are modestly lower across all values of $\lambda$, while the overall number of validation-successful runs is the same: under the validation classification defined in Section~\ref{subsubsec:4eval}, 66 out of 70 runs satisfy $r^\mathrm{val}\leq10$ in both cases. 
To examine the effect of loss balancing beyond the overall success counts, we next compare the family-level validation performance in greater detail. 
Detailed unweighted training losses and validation-successful run counts are reported in Section~S1.2 of the Supplementary Material.

\begin{figure}[tbp] 
    \centering 
    \includegraphics[width=0.75\textwidth]{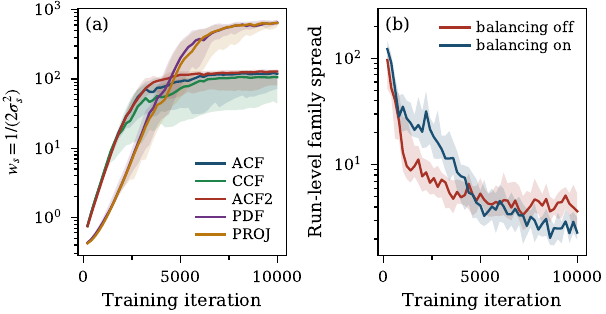} 
    \caption{Self-adaptive loss balancing for the coupled Duffing benchmark at $\lambda=4$. 
    (a) Evolution of the adaptive weights across 20 runs with loss balancing enabled. 
    Lines show medians, and shaded bands indicate interquartile ranges. 
    The median PDF and PROJ curves largely overlap during the later stages of training. 
    (b) Median run-level family spread with and without loss balancing at each validation step, computed using Eq.~\eqref{eq:family_spread_updated}. 
    Shaded bands indicate interquartile ranges across runs.}
\label{fig:4lb_updated} 
\end{figure}

We demonstrate how the adaptive weights evolve during optimization using the 20 balancing-enabled runs at $\lambda=4$. 
Figure~\ref{fig:4lb_updated}(a) shows the evolution of the adaptive weights $w_s=1/(2\sigma_s^2)$ associated with the individual statistical loss families. 
The weights generally increase during training and become approximately steady at later iterations. 
The three time-correlation families, ACF, CCF, and ACF2, exhibit similar weight evolution and approach values of comparable magnitude, while the two probability-density families, PDF and PROJ, likewise exhibit similar behavior and approach substantially larger values. 
This separation is consistent with differences in the optimization sensitivities and convergence behaviors of the loss families associated with distinct types of statistical quantities. 
The adaptive weights adjust their relative contributions to the training objective as optimization proceeds.

We next examine whether this adaptive reweighting results in more balanced reference-normalized validation performance across the loss families. 
To quantify this balance within each run, we define the spread among the family-level validation-loss ratios at each validation step as 
\begin{equation} 
    \mathrm{spread} = \frac{\max_s r_s^{\mathrm{val}}}{\min_s r_s^{\mathrm{val}}},
\label{eq:family_spread_updated} 
\end{equation} 
where the maximum and minimum are taken over the applicable statistical families within the same run. 
Values closer to one indicate more uniform family-level validation-loss ratios within that run. 
Figure~\ref{fig:4lb_updated}(b) compares the median run-level spread with and without loss balancing. 
Although the balancing-enabled runs initially exhibit a larger spread, the spread decreases as optimization proceeds and eventually becomes substantially smaller than that observed without balancing.

This behavior suggests that the adaptive reweighting progressively improves the balance of reference-normalized validation performance across the loss families.
At the retained checkpoints, loss balancing lowers the median $r^\mathrm{val}$ at every coupling strength.
It also lowers the median run-level spread at all nonzero coupling strengths, while the two configurations yield similar spread values at $\lambda=0$.
Detailed family-level validation-loss ratios and spread values across all coupling strengths are reported in Section~S1.2 of the Supplementary Material.

The threshold $r^\mathrm{val}=10$ provides a practical distinction between validation-unsuccessful runs and runs that do not meet the failure criterion, but it does not distinguish among the latter according to their reference-normalized validation performance. 
To assess the effect of loss balancing across different levels of validation performance, we therefore examine a range of thresholds for $r^\mathrm{val}$. 
For a prescribed threshold $r$, we consider whether and when each run attains $r^\mathrm{val}\leq r$ during training. 
Smaller values of $r$ represent increasingly stringent levels of agreement relative to the finite-sample reference level.

\begin{figure}[tbp]
    \centering
    \includegraphics[width=0.75\textwidth]{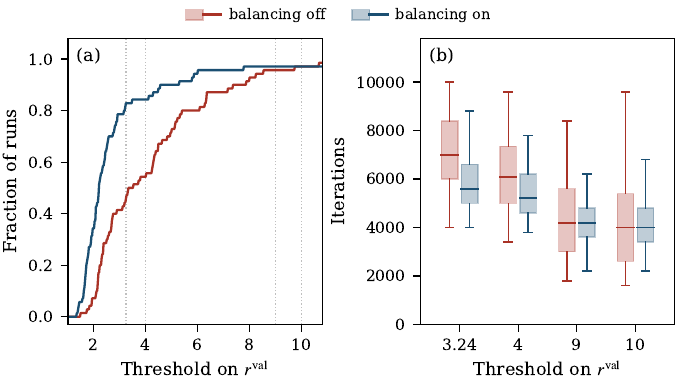}
    \caption{Validation-loss-ratio threshold attainment over the full training history, pooled across all six coupling strengths.
    (a) Fraction of runs reaching each threshold at any validation step.
    (b) First-attainment iterations among the runs reaching each indicated threshold. 
    Boxes show the interquartile range, with the median marked inside; whiskers extend to the most extreme observations within $1.5$ times the interquartile range.
    Outliers are not shown.}
    \label{fig:4lb_attainment}
\end{figure}

Figure~\ref{fig:4lb_attainment} summarizes this threshold-attainment analysis over the full validation histories, pooling all 70 runs for each loss-balancing configuration across the six coupling strengths. 
Panel~(a) reports the fraction of runs that attain each threshold at any validation step. 
The advantage of loss balancing becomes more apparent as the threshold is tightened, with a larger fraction of balancing-enabled runs attaining the prescribed thresholds. 
Panel~(b) reports the iteration at which each threshold is first attained among the runs that reach that threshold. 
At tighter thresholds, loss balancing also tends to yield smaller median first-attainment iterations, whereas the distributions overlap substantially at $r=10$. 
Because panel~(b) includes only runs that reach each threshold, the first-attainment iterations should be interpreted together with the fractions of runs reaching that threshold in panel~(a).

We finally assess the computational overhead introduced by loss balancing. 
Per-iteration cost is measured in a standalone benchmark rather than from the training logs. 
For each setting, we time 30 repetitions of a complete training iteration, including the forward rollout, family-loss evaluation, backward pass, and optimizer updates, after five warm-up repetitions, with CUDA synchronization around each repetition. 
The input noise tensor is drawn once and reused across repetitions, and periodic validation and checkpointing are not included. 
The median iteration time increases from $1.05\times 10^{-2}$~s without loss balancing to $1.13\times 10^{-2}$~s with loss balancing, corresponding to an increase of $7\%$.

Overall, these results show that self-adaptive loss balancing improves the balance of reference-normalized validation performance across the statistical families and allows a larger fraction of runs to satisfy more stringent validation-loss-ratio thresholds, while introducing only a modest increase in per-iteration computational cost.

\subsection{Projected-PDF Constraints and Joint-Distribution Accuracy}
\label{subsec:4proj}

We next assess the role of the projected-PDF constraints introduced in Section~\ref{subsec:projected_pdf} in learning the joint distribution of the multidimensional target process. 
Throughout this subsection, we fix the white-noise input dimension at $d_\mathrm{in}=2$ and enable self-adaptive loss balancing, while varying the number of projection frames over $n_\mathrm{f}\in\{1,2,3,4\}$.
The case $n_\mathrm{f}=1$ contains only the coordinate frame and therefore introduces no additional projected-PDF loss terms, whereas $n_\mathrm{f}=4$ corresponds to the full set of projection frames defined in Eq.~\eqref{eq:frames2d}.
For intermediate values of $n_\mathrm{f}$, the non-coordinate frames are added sequentially in the order given in Eq.~\eqref{eq:frames2d}.

We assess joint-distribution accuracy using the relative joint-density error $e_\mathrm{JOINT}$.
The joint density of the SINN-generated trajectories is estimated using a bivariate KDE on the product grid $[-5,5]^2$, with the bandwidth specified in Table~\ref{tab:4params}.
The reference density is obtained from Eq.~\eqref{eq:gibbs} by numerical quadrature, transformed to standardized coordinates, and convolved with the same bivariate Gaussian kernel used in the KDE. 
We define $e_\mathrm{JOINT}$ as the relative $L_2$ discrepancy between the estimated and reference densities evaluated on the common grid.
Importantly, $e_\mathrm{JOINT}$ enters neither the training objective nor the checkpoint-selection criterion and therefore provides an independent measure of joint-distribution accuracy.

\begin{figure}[tbp]
    \centering
    \includegraphics[width=0.6\textwidth]{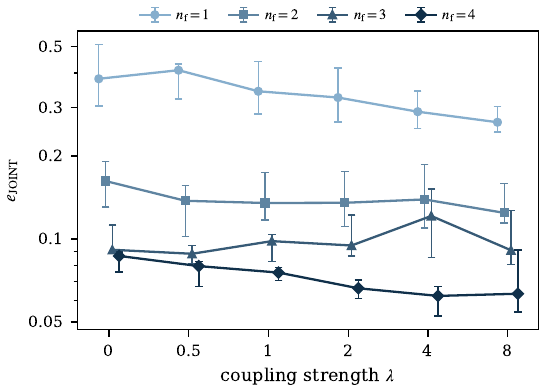}
    \caption{Relative joint-density error $e_\mathrm{JOINT}$ for different numbers of projection frames, at fixed $d_\mathrm{in}=2$ with loss balancing enabled. 
    Markers show medians over runs, and error bars indicate interquartile ranges.
    Markers for different $n_\mathrm{f}$ values at each coupling strength are slightly offset horizontally for visual clarity.
    The vertical axis is logarithmic.}
\label{fig:4joint_sweep}
\end{figure}

Figure~\ref{fig:4joint_sweep} shows that the median $e_\mathrm{JOINT}$ decreases monotonically as the number of projection frames increases at every coupling strength. 
Adding the first non-coordinate frame substantially reduces the error relative to the coordinate-only configuration, and additional frames yield further improvements.
These results show that adding projected-PDF constraints improves the reproduction of the joint distribution, even though the joint density itself is not included in the training objective.

\begin{figure}[tbp]
    \centering
    \includegraphics[width=\textwidth]{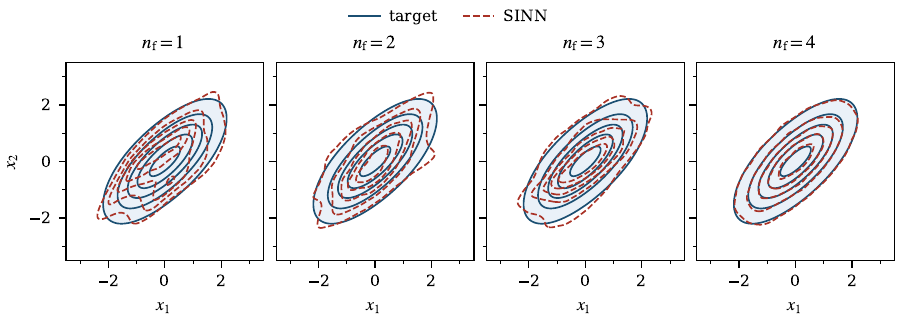}
    \caption{Reference joint density (solid) and joint density estimated from the SINN-generated trajectories (dashed) at $\lambda=4$ for $n_\mathrm{f} = 1,2,3,4$. 
    Contour levels are chosen to enclose $10$, $30$, $50$, $70$, and $90$ percent of the reference probability mass and are also used for the estimated density.
    Each panel shows one of the two middle-ranked runs according to the retained-checkpoint validation score.}
\label{fig:4joint}
\end{figure}

To examine the improvement in joint-distribution accuracy more directly, we compare the estimated and reference joint densities at $\lambda=4$. 
Figure~\ref{fig:4joint} shows that the coordinate-only configuration exhibits substantial discrepancy from the reference density, whereas adding non-coordinate projection frames brings the estimated density contours into progressively closer agreement with the reference. 
To further assess whether this improvement is specific to the projection directions used during training, we evaluate four additional held-out directions. 
For these directions, the median relative density error also decreases monotonically as the number of projection frames increases at every coupling strength (see Section~S1.3 of the Supplementary Material).
Together, these results demonstrate that the projected-PDF constraints progressively improve joint-distribution accuracy and that the improvement extends to projection directions not explicitly used during training.

\subsection{White-Noise Input Dimension} 
\label{subsec:4din} 

To assess the effect of the white-noise input dimension, we compare $d_\mathrm{in}=1$, $2$, and $3$ while fixing $n_\mathrm{f}=4$ and enabling self-adaptive loss balancing. 
To isolate the effect of the number of active noise channels, we use a fixed four-channel LSTM input in all configurations, with independent white noise supplied to the first $d_\mathrm{in}$ channels and zeros supplied to the remaining channels. 
This construction keeps the network architecture and parameter count unchanged while varying only the dimensionality of the stochastic input.

\begin{figure}[tbp] 
    \centering 
    \includegraphics[width=0.6\textwidth]{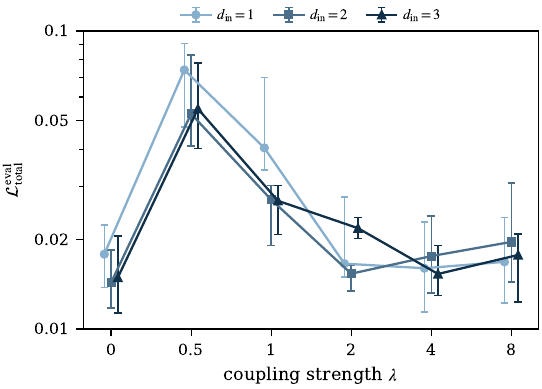} 
    \caption{Total evaluation loss $\mathcal{L}_\mathrm{total}^\mathrm{eval}$ for $d_\mathrm{in}=1,2,3$ at fixed $n_\mathrm{f}=4$ with loss balancing enabled. 
    Markers show medians over runs, and error bars indicate interquartile ranges.  
    Coupling strengths are equally spaced for display, and markers for different $d_\mathrm{in}$ values are slightly offset horizontally for visual clarity. 
    The vertical axis is logarithmic.}
\label{fig:4din} 
\end{figure}

We assess performance using the total evaluation loss $\mathcal{L}_\mathrm{total}^\mathrm{eval}$ defined in Eq.~\eqref{eq:total_evaluation_loss}. 
As described in Section~\ref{subsubsec:4eval}, the correlation-based families are evaluated against statistics estimated from the final-evaluation ensemble $\mathcal{T}$, whereas the PDF and PROJ families are evaluated against the quadrature-derived reference densities. 
Figure~\ref{fig:4din} compares the evaluation losses across the six coupling strengths. 
Increasing the input dimension from $d_\mathrm{in}=1$ to $2$ reduces the median evaluation loss at $\lambda=0$, $0.5$, and $1$, whereas no consistent improvement is observed at the larger coupling strengths. 
Increasing the input dimension further to $d_\mathrm{in}=3$ likewise provides no consistent improvement over $d_\mathrm{in}=2$.

We note that the improvement from $d_\mathrm{in}=1$ to $2$ at $\lambda=0$ is consistent with the structure of the uncoupled system: the two oscillators are independent and driven by independent stochastic forcing, making two independent noise channels a natural input representation. 
We also note that the loss magnitudes across different coupling strengths should not be compared directly. 
In particular, the relatively large total evaluation losses at weak nonzero coupling may be influenced by the CCF family, since the small magnitude of the target CCF can amplify modest absolute discrepancies through the relative normalization.
We therefore interpret the effect of $d_\mathrm{in}$ through comparisons performed at each fixed coupling strength.

The validation-successful run counts provide a consistent overall picture. 
Under the validation classification defined in Section~\ref{subsubsec:4eval}, $61/70$ runs with $d_\mathrm{in}=1$ satisfy $r^\mathrm{val}\leq10$, whereas $66/70$ runs satisfy this condition for both $d_\mathrm{in}=2$ and $d_\mathrm{in}=3$. 
Detailed evaluation losses and validation-successful run counts for the individual coupling strengths are reported in Section~S1.4 of the Supplementary Material. 
Thus, increasing the input dimension from one to two channels improves performance at several coupling strengths and increases the overall number of validation-successful runs, whereas adding a third channel provides no consistent additional benefit.
These results support $d_\mathrm{in}=d=2$ as a practical choice for this two-dimensional benchmark.

\section{Applications}
\label{sec5}

Having validated the individual methodological components of multidimensional SINN using the controlled benchmark in Section~\ref{sec4}, we now apply the framework to low-dimensional observables derived from much higher-dimensional stochastic systems. 
These applications examine the use of multidimensional SINN as a coarse stochastic surrogate that directly generates trajectories of selected observables without resolving the underlying microscopic degrees of freedom.
For the Fermi--Pasta--Ulam (FPU) chain in Section~\ref{subsec5.1}, we consider two- and three-component observables formed from spatially separated particle positions.
For the lattice Lotka--Volterra (LLV) model in Section~\ref{subsec5.2}, we consider coarse population densities obtained by aggregating the microscopic lattice dynamics generated by kinetic Monte Carlo simulations.

\subsection{Fermi--Pasta--Ulam Chain}
\label{subsec5.1}

Low-dimensional observables of an interacting particle system generally do not obey a closed Markovian equation because the unresolved degrees of freedom can introduce memory and effective stochastic forcing. 
The original SINN study considered a single coarse position in a Fermi--Pasta--Ulam (FPU) chain~\cite{ZhuTangKim2023}.
Here, we extend this setting to two- and three-component processes formed from multiple spatially separated positions and examine whether multidimensional SINN can reproduce their joint distributional and temporal statistics.

We consider the Langevin dynamics of a 100-particle FPU chain. 
With periodic indexing, the bond stretches $r_j$ and particle momenta $p_j$ ($j=1,\ldots,100$) evolve as~\cite{ZhuLei2022, ZhuTangKim2023}
\begin{equation} 
\begin{aligned} 
    \dot{r}_j &= p_j-p_{j-1},\\ 
    \dot{p}_j &= V'(r_{j+1})-V'(r_j)-\gamma p_j +\sqrt{2\gamma/\beta}\,\xi_j(t),
\end{aligned} 
\label{eq:fpu_chain} 
\end{equation}
where $\xi_j(t)$ are independent standard Gaussian white noises.
The bond interaction is described by the same quadratic--quartic form used for the individual Duffing oscillators in Section~\ref{subsec:4benchmark},
\begin{equation} 
    V(r)=\frac{\alpha}{2}r^2+\frac{\kappa}{4}r^4.
\end{equation}
We set the potential parameters to $\alpha=\kappa=1$, the inverse temperature to $\beta=1$, and the damping coefficient to $\gamma=1$. 
Taking $q_0=0$ as the reference position, the particle positions are recovered from the bond stretches as $q_j=\sum_{k=1}^j r_k$, with $q_{100}=q_0$ under the periodic constraint.

Target trajectories are generated by integrating Eq.~\eqref{eq:fpu_chain} using the Euler--Maruyama method with time step $dt=10^{-2}$. 
After discarding an initial transient of 250 time units, states are retained at a sampling interval of $\Delta t=0.5$, and the first $T=400$ retained states of each trajectory define the analysis window.
We construct the two- and three-dimensional target processes from spatially separated particle positions: 
\begin{equation} 
    \boldsymbol{x}^\mathrm{2D}_t =\bigl(\widetilde{q}_{25}(t),\widetilde{q}_{75}(t)\bigr), \quad
    \boldsymbol{x}^\mathrm{3D}_t =\bigl(\widetilde{q}_{25}(t),\widetilde{q}_{50}(t), \widetilde{q}_{75}(t)\bigr), 
\label{eq:fpu_targets}
\end{equation} 
where $\widetilde{q}_j$ denotes the standardized position of particle $j$. 
Because the equilibrium mean positions are zero, standardization involves only scaling, with the component scales estimated from the post-transient trajectories in the pre-generated FPU archive. 
The two targets share the spatially separated outer components $\widetilde q_{25}$ and $\widetilde q_{75}$, while the three-dimensional target additionally includes the intervening component $\widetilde q_{50}$.
Further details of the archive and preprocessing are provided in Section~S2.1 of the Supplementary Material.

The FPU experiments follow the general training, validation, and evaluation framework of Sections~\ref{sec3} and~\ref{sec4}, with the following application-specific settings. 
We use a fitting ensemble of $N=400$ trajectories, nine reference blocks of the same size, a validation ensemble of 4,000 trajectories, and an independently generated final-evaluation ensemble of 12,000 trajectories. 
The FPU models use three stacked LSTM layers with $d_\mathrm{in}=d$ ($d=2,3$) and are trained for $2\times10^4$ iterations. 
For the FPU experiments, we use the absolute loss for all five statistical families and optimize the positive scale parameters $\sigma_s$ directly.
For each run, we retain the checkpoint that minimizes the unweighted sum of the five absolute validation losses and subsequently classify the retained model using the validation classification defined in Section~\ref{subsubsec:4eval}. 
The final-evaluation ensemble $\mathcal{T}$ is not used for either checkpoint selection or validation classification. 
Additional preprocessing and optimization details, including the treatment of the scale parameters, are provided in Section~S2.1 of the Supplementary Material.

Before assessing the learned multidimensional statistics, we perform a modest network-width sweep to select a common model capacity for the two target processes. 
We vary the hidden width among 30, 50, 75, and 110 units per LSTM layer while keeping the remaining model and training settings fixed. 
A width of 75 provides reliable validation performance for both targets, with 16 of 20 two-dimensional runs and 18 of 20 three-dimensional runs satisfying $r^\mathrm{val}\leq10$. 
Increasing the width to 110 does not increase the fraction of validation-successful runs for either target.
We therefore use a width of 75 for the analyses below. 
The complete width sweep is reported in Section~S2.2 of the Supplementary Material.

For the final evaluation of the two- and three-dimensional cases, we generate a new ensemble of $N=400$ SINN trajectories from each of the 20 retained models using newly sampled white-noise input sequences. 
The models retain their classification as validation-successful or validation-unsuccessful according to the procedure in Section~\ref{subsubsec:4eval}.
The five statistical families are evaluated against the corresponding target statistics from the independent final-evaluation ensemble $\mathcal{T}$ using the same absolute-loss definitions employed during validation. 
For each family $s$, the final-evaluation ratio is defined as 
\begin{equation} 
    r_s^\mathrm{eval} = \frac{\mathcal{L}_s^\mathrm{eval}}{\mathcal{L}_s^\mathrm{ref,eval}},
\end{equation} 
where $\mathcal{L}_s^\mathrm{ref,eval}$ is the median absolute discrepancy obtained by comparing the nine reference blocks with the same final-evaluation target. 
Because both the evaluation and reference discrepancies use the same absolute-loss definition, their ratio measures the SINN discrepancy relative to the finite-sample variability represented by independent target ensembles.
As in the validation analysis, values of order unity indicate discrepancies comparable to the finite-sample variability represented by the reference blocks.

\begin{figure*}[tbp]
    \centering
    \includegraphics[width=\textwidth]{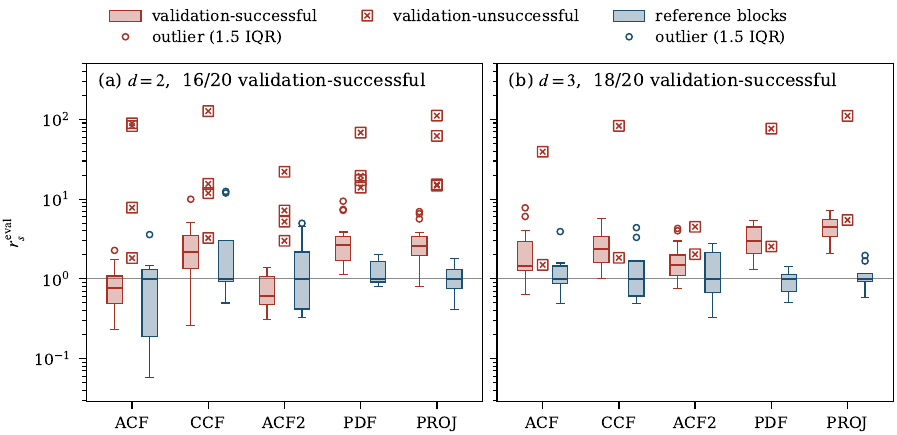}
    \caption{Final-evaluation family-loss ratios for the FPU models with width 75: 
    (a) two-dimensional and (b) three-dimensional cases. 
    Red boxes summarize validation-successful runs, while crossed open squares indicate validation-unsuccessful runs. 
    Blue boxes show the corresponding ratios computed from the nine reference blocks. 
    Circles denote boxplot outliers beyond $1.5$ times the interquartile range. 
    The horizontal line indicates $r_s^\mathrm{eval}=1$.}
\label{fig:fpu_family_ratios}
\end{figure*}

Figure~\ref{fig:fpu_family_ratios} compares the resulting family-level ratios according to validation outcome: validation-successful runs are summarized collectively, while validation-unsuccessful runs are displayed individually.
Detailed numerical results for the final evaluation are reported in Section~S2.3 of the Supplementary Material.
Among the validation-successful runs, the median family-level ratios range from $0.61$ to $2.64$ for the two-dimensional target and from $1.45$ to $4.44$ for the three-dimensional target. 
For the three-dimensional target, the largest median ratios occur for the PDF and PROJ families, indicating that the distributional statistics remain the most challenging among the prescribed statistical objectives. 
In contrast, the validation-unsuccessful runs exhibit substantially larger final-evaluation ratios in one or more statistical families.
Thus, the distinction between validation-successful and validation-unsuccessful models generally persists under independent final evaluation, supporting the use of the validation classification as a practical means of identifying models that avoid substantial disagreement with the target statistics.

\begin{figure*}[tbp]
    \centering
    \includegraphics[width=\textwidth]{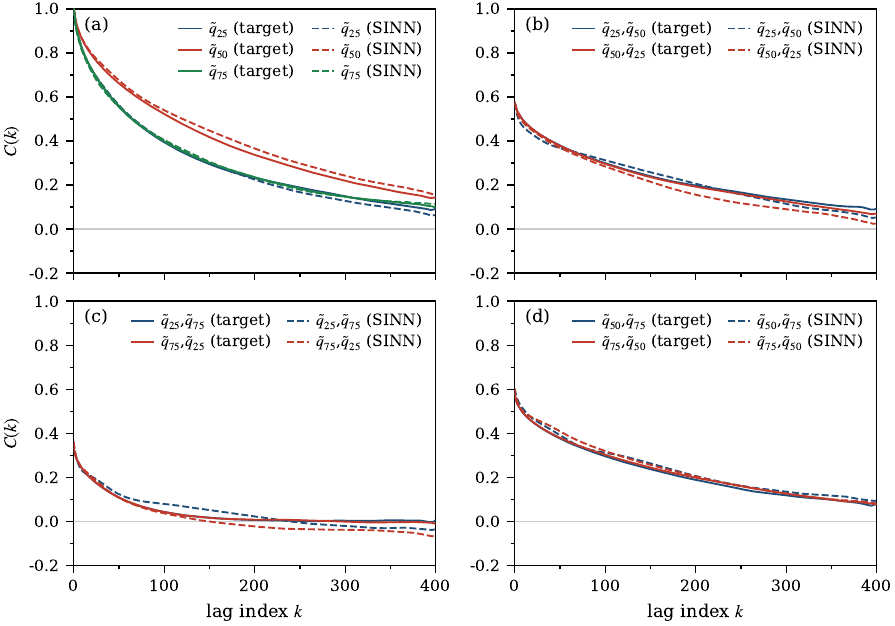}
    \caption{Temporal correlations for a representative validation-successful three-dimensional FPU model. 
    (a) Component ACFs and (b)--(d) ordered CCFs for the three component pairs. 
    Solid curves show correlations computed from the 12,000 target trajectories in the final-evaluation ensemble $\mathcal{T}$, and dashed curves show the corresponding results computed from the 12,000-trajectory SINN ensemble.
    Colors distinguish the components in panel (a) and the two orderings of each component pair in panels (b)--(d). 
    All panels use the same vertical scale to facilitate comparison of the correlation magnitudes and target--SINN discrepancies across the component pairs.    
    For this diagnostic comparison, the correlations are normalized at each lag by the number of contributing sample pairs, $N(T-k)$, rather than by the fixed factor $NT$ used in the statistical loss evaluation.}
\label{fig:fpu_corr}
\end{figure*}

\begin{figure*}[tbp]
    \centering
    \includegraphics[width=\textwidth]{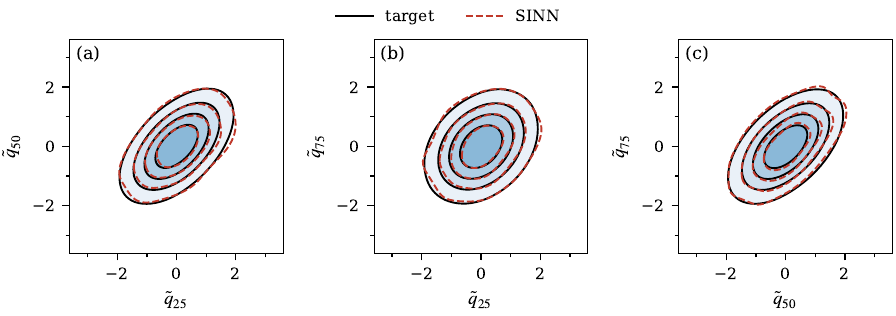}
    \caption{Pairwise joint densities for the same representative three-dimensional FPU model as in Figure~\ref{fig:fpu_corr}:  
    (a) $(\widetilde{q}_{25},\widetilde{q}_{50})$, (b) $(\widetilde{q}_{25},\widetilde{q}_{75})$, and (c) $(\widetilde{q}_{50},\widetilde{q}_{75})$. 
    The densities are estimated from the 12,000 target trajectories in the final-evaluation ensemble $\mathcal{T}$ and the 12,000-trajectory SINN ensemble.
    Filled regions and solid contours show the target densities, while dashed contours show the corresponding SINN densities. 
    Contour levels are $0.2$, $0.4$, $0.6$, and $0.8$ times the maximum target density in each panel.}
\label{fig:fpu_joint}
\end{figure*}

For the detailed diagnostics in Figures~\ref{fig:fpu_corr} and~\ref{fig:fpu_joint}, we select a representative validation-successful three-dimensional model.
For this model, we generate a separate ensemble of 12,000 SINN trajectories using newly sampled white-noise inputs and compare the resulting statistics with those computed from the 12,000 target trajectories in the independent final-evaluation ensemble $\mathcal{T}$.

While the family-level ratios provide an aggregate measure of statistical accuracy, they do not reveal how well the detailed temporal-correlation structure is reproduced.
Figure~\ref{fig:fpu_corr} therefore compares the component ACFs and ordered CCFs for this representative model.
The component ACFs reproduce the overall decay behavior of the target correlations, and the CCFs involving the intervening component $\widetilde q_{50}$ likewise capture their overall magnitudes and decay patterns.
The most noticeable discrepancies occur for the weaker correlations between the maximally separated outer components $\widetilde{q}_{25}$ and $\widetilde q_{75}$.
In particular, the SINN estimate of the ordered correlation $C_{\widetilde{q}_{75},\widetilde{q}_{25}}$ decays more rapidly than the target and becomes negative at intermediate lags, while the opposite ordering exhibits a more modest discrepancy. 
However, these outer-component correlations are substantially weaker than the ACFs and the CCFs involving $\widetilde q_{50}$, so the corresponding absolute discrepancies remain relatively modest when viewed on a common correlation scale.

We further assess the cross-component dependence captured by the same representative three-dimensional model using the three pairwise joint densities shown in Figure~\ref{fig:fpu_joint}. 
The densities are estimated from the same 12,000-trajectory SINN and target ensembles used for the temporal-correlation diagnostics in Figure~\ref{fig:fpu_corr}. 
The SINN-generated contours reproduce the principal locations, orientations, and spreads of the target distributions, with modest discrepancies primarily in some of the outer contours. 
These bivariate densities are included neither directly in the training objective nor in the checkpoint-selection criterion, although the projected-PDF family incorporates information about the joint distribution through one-dimensional projections.
Their agreement therefore provides an additional diagnostic that the multidimensional SINN captures the pairwise dependence among the three coarse FPU observables.

Taken together, the across-run evaluation shows that validation-successful multidimensional SINN models reproduce the principal prescribed statistics of both target processes, while the representative three-dimensional diagnostics illustrate the corresponding temporal and pairwise distributional agreement in detail. 
These results support the use of multidimensional SINN as a coarse stochastic surrogate for multiple spatially separated FPU observables, although reproducing the weak temporal correlations between the most widely separated components remains comparatively challenging.

\subsection{Lattice Lotka--Volterra Model}
\label{subsec5.2}

We next consider a lattice Lotka--Volterra (LLV) model~\cite{KimRickeVanVoorhis2021} and construct coarse-grained prey and predator densities from simulations of its microscopic lattice dynamics. 
We examine whether multidimensional SINN can reproduce the joint distributional and temporal statistics of these coarse variables. 
In addition, we assess whether the learned statistical properties are maintained as generation continues beyond the retained trajectory length used for training.

\begin{figure*}[tbp] 
    \centering 
    \includegraphics[width=\textwidth]{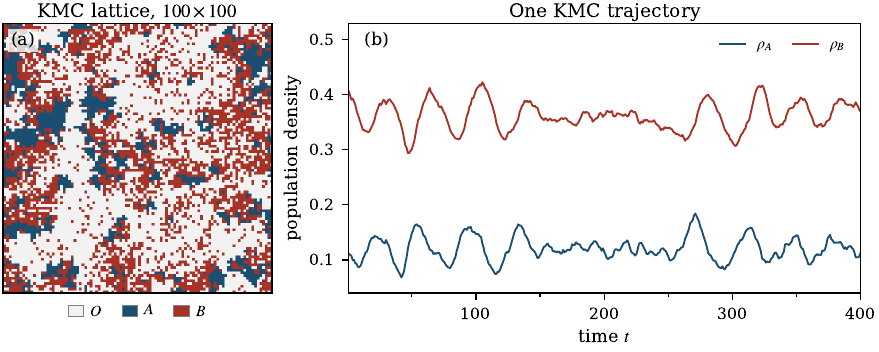} 
    \caption{Lattice Lotka--Volterra target data.
    (a) A post-transient snapshot of the KMC lattice.
    (b) Prey and predator densities $\rho_A$ and $\rho_B$ from a separate KMC trajectory, plotted in physical time.}
\label{fig:llv_setup}
\end{figure*}

The microscopic lattice dynamics are simulated using kinetic Monte Carlo (KMC)~\cite{AndersenPanosettiReuter2019}.
Each site of a periodic $100\times100$ square lattice is either vacant ($O$), occupied by prey ($A$), or occupied by predators ($B$). 
The dynamics are governed by the reactions 
\begin{equation} 
    O+A \xrightarrow{k_1} A+A,\quad 
    A+B \xrightarrow{k_2} B+B,\quad 
    B \xrightarrow{k_3} O,
\label{eq:llv_reactions}
\end{equation}
where the first two reactions act on nearest-neighbor pairs of lattice sites, while the third occurs at an individual site.
We use $k_1=0.5$, $k_2=0.3$, and $k_3=0.1$. 
With $N_\mathrm{site}=10^4$ lattice sites, we define the two-dimensional target process as 
\begin{equation} 
    \boldsymbol{X}_t=(\rho_A(t),\rho_B(t)),
\label{eq:llv_density}
\end{equation}
where $\rho_A(t)$ and $\rho_B(t)$ denote the fractions of lattice sites occupied by prey and predators, respectively; the remaining fraction $\rho_O(t)=1-\rho_A(t)-\rho_B(t)$ corresponds to vacant sites.
Figure~\ref{fig:llv_setup} illustrates a KMC lattice configuration and the corresponding coarse population densities.

Target trajectories are sampled at intervals of $\Delta t=1$ after discarding an initial transient of 400 physical time units. 
We use a fitting ensemble of $N=400$ trajectories, nine reference blocks of the same size, a validation ensemble of 4,000 trajectories, and a final-evaluation ensemble of 12,000 trajectories, with the first $T=400$ retained states of each trajectory defining the analysis window. 
All target statistics are estimated empirically from the KMC trajectories. 
The standardization constants and projection scales are estimated from the fitting ensemble, while the density grids and bandwidths are fixed throughout training, validation, and evaluation. 
Further details of the target-data archive, transient assessment, and numerical settings are provided in Section~S3.1 of the Supplementary Material.

For the LLV application, we follow the general training configuration of Section~\ref{subsubsec:4train}, using two stacked LSTM layers with ten hidden units per layer, a white-noise input dimension of $d_\mathrm{in}=2$, and self-adaptive loss balancing. 
All five statistical families use the relative loss, and each model is trained for $10^4$ iterations using the Adam optimizer, following the warm-up and indexing convention of Section~\ref{subsec:optim_strat}. 
For each run, we retain the checkpoint that minimizes the unweighted sum of the five validation losses, as in Section~\ref{subsubsec:4eval}. 
The retained model is then evaluated using a fresh independent ensemble of $N=400$ generated trajectories to compute the validation ratios. 
It is classified as validation-successful or validation-unsuccessful according to whether $r^\mathrm{val}\leq10$ or $r^\mathrm{val}>10$, respectively. 
For final evaluation, each retained model generates an independent ensemble of $N=400$ trajectories with 800 retained time steps. 
The first 400 retained steps, $t=1,\ldots,400$, define the initial segment corresponding to the retained window used during training, while the subsequent 400 steps, $t=401,\ldots,800$, define the continuation segment. 
The recurrent state evolves continuously across the two segments without being reset. 
Because the target process is stationary, the continuation segment provides a test of whether the learned statistical properties persist beyond the trajectory window on which the training statistics are evaluated.

\begin{figure*}[tbp]
    \centering
    \includegraphics[width=\textwidth]{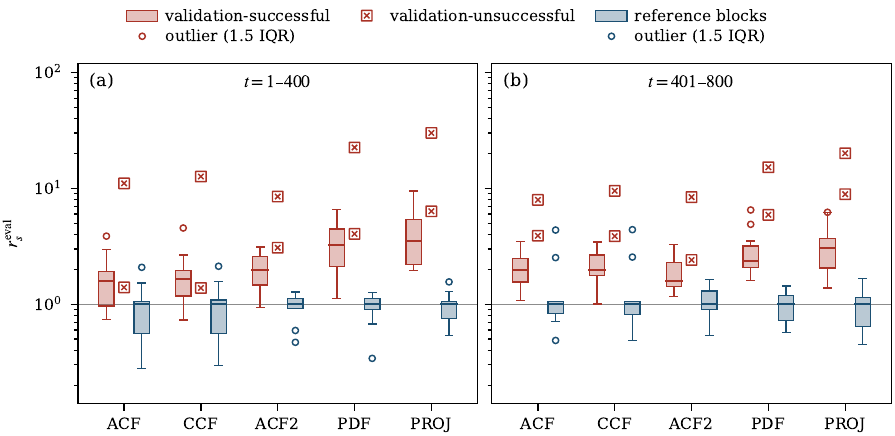}
    \caption{LLV final-evaluation family-loss ratios for (a) the initial segment ($t=1$--$400$) and (b) the continuation segment ($t=401$--$800$). 
    Red boxes summarize the 18 validation-successful runs, while crossed open squares indicate the two validation-unsuccessful runs. 
    Blue boxes summarize the corresponding ratios for the nine reference blocks. 
    Generated and reference-block ensembles each contain 400 trajectories and are compared with the corresponding segment of the 12,000-trajectory final-evaluation ensemble. 
    Family-loss ratios are normalized using segment-specific reference discrepancies.}
\label{fig:llv_ratios}
\end{figure*}

Of the 20 independent runs, 18 are classified as validation-successful using the threshold $r^\mathrm{val}\leq10$ defined in Section~\ref{subsubsec:4eval}. 
Final-evaluation family ratios are defined analogously to the validation ratios in Section~\ref{subsubsec:4eval}, using the corresponding segment of the final-evaluation ensemble as the target and the associated segment-specific reference discrepancies. 
Figure~\ref{fig:llv_ratios} compares these ratios for the initial and continuation segments. 
For both segments, the correlation-based families are generally closer to the finite-sample reference level than the PDF and PROJ families. 
Because the ratios for the two segments use segment-specific reference discrepancies, they should not be compared directly to assess changes between the segments. 
A paired comparison using a common reference scale shows no consistent direction of change across the five statistical families, providing no evidence of systematic degradation from the initial to the continuation segment; details are provided in Section~S3.2 of the Supplementary Material.

We next examine the temporal and distributional statistics of a representative validation-successful model selected based on the retained-checkpoint validation score. 
For this model, we generate a separate ensemble of 12,000 SINN trajectories using newly sampled white-noise inputs and use this ensemble for the detailed diagnostics below. 
We focus these diagnostics on the continuation segment because it lies beyond the trajectory window on which the training statistics are evaluated. 
For a stationary target process, agreement over this segment therefore provides a direct test of whether the learned statistical behavior persists as generation continues.

\begin{figure*}[tbp]
    \centering
    \includegraphics[width=\textwidth]{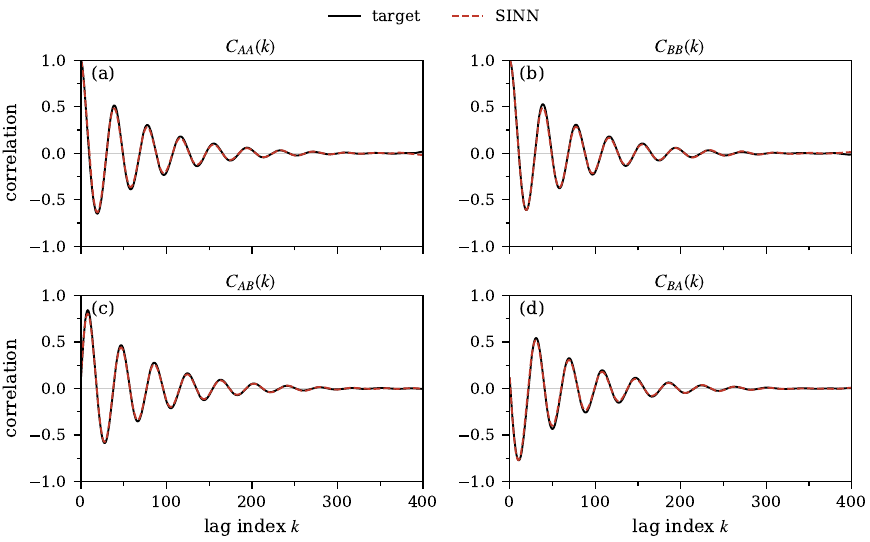}
    \caption{Temporal correlations over the continuation segment ($t=401$--$800$) for the representative LLV model. 
    Black solid curves show correlations computed from the 12,000-trajectory KMC target ensemble, and red dashed curves show the corresponding results computed from the 12,000-trajectory SINN ensemble. 
    For this diagnostic comparison, the correlations are normalized at each lag by the number of contributing sample pairs, $N(T-k)$, rather than by the fixed factor $NT$ used in the statistical loss evaluation.}
\label{fig:llv_corr}
\end{figure*}

Figure~\ref{fig:llv_corr} examines the temporal correlations for the representative model.
Over the continuation segment, the generated ACFs and CCFs closely reproduce the oscillatory structure of the target correlations, including their dominant periods, amplitudes, and decay behavior. 
The overall quality of agreement is comparable between the initial and continuation segments, although the latter exhibits slightly larger discrepancies. 
The corresponding initial-segment comparison is provided in Section~S3.2 of the Supplementary Material.

\begin{figure*}[tbp]
    \centering
    \includegraphics[width=\textwidth]{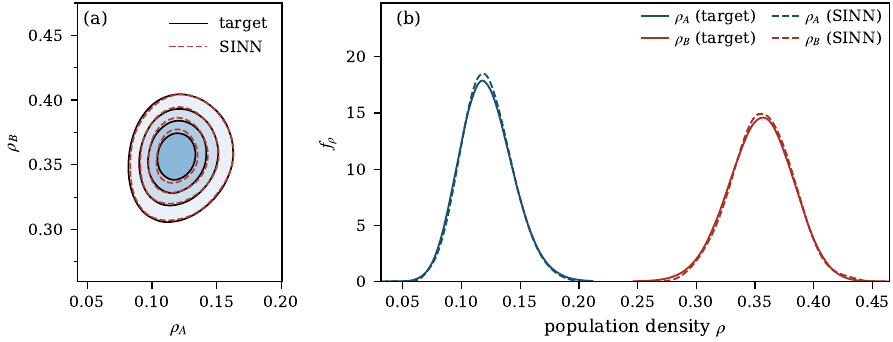}
    \caption{Joint and marginal densities over the continuation segment ($t=401$--$800$) for the same representative LLV model, computed using 12,000 target and SINN trajectories. 
    (a) Target joint density (filled regions and black solid contours) and SINN joint density (red dashed contours). 
    Contour levels are $0.2$, $0.4$, $0.6$, and $0.8$ times the maximum target density. 
    (b) Marginal densities of $\rho_A$ and $\rho_B$; solid and dashed curves show the target and SINN results, respectively.}
\label{fig:llv_density}
\end{figure*}

Figure~\ref{fig:llv_density} further examines the distributional statistics of the same representative model. 
Over the continuation segment, the generated joint density reproduces the principal location, shape, and orientation of the KMC target distribution, with modest discrepancies primarily in the outer contours. 
The marginal densities of $\rho_A$ and $\rho_B$ also show close agreement with the corresponding target densities. 
The overall quality of distributional agreement is comparable between the initial and continuation segments, with the corresponding initial-segment comparison provided in Section~S3.2 of the Supplementary Material. 
The bivariate density is included neither directly in the training objective nor in the checkpoint-selection criterion, although the projected-PDF family incorporates information about the joint distribution through one-dimensional projections. 
The agreement of the bivariate density therefore provides an additional diagnostic of the cross-component dependence captured by SINN.

All post-warm-up outputs from the 20 retained models remain finite and satisfy the physical constraints $\rho_A\geq0$, $\rho_B\geq0$, and $\rho_A+\rho_B\leq1$. 
Although the initial and continuation segments exhibit comparable statistical agreement, even small differences between these finite windows could in principle reflect a slow drift that accumulates over much longer generation times. 
To examine this possibility, we perform a separate $10^6$-step stress test of the representative model. 
The generated process remains numerically stable and physically admissible, with no evidence of substantial statistical drift over the much longer generation interval; details are provided in Section~S3.3 of the Supplementary Material. 
Overall, these results show that multidimensional SINN reproduces the principal temporal and distributional features of the coupled prey--predator density process, maintains comparable statistical agreement beyond the trajectory window on which the training statistics are evaluated, and shows no evidence of progressive drift over much longer generation times.

\section{Conclusion}
\label{sec6}

In this work, we extended the statistics-informed neural network (SINN) framework from scalar to multidimensional stochastic processes. 
The main challenge in this extension is not merely to increase the output dimension, but to represent cross-component dependence while balancing multiple statistical objectives with different optimization sensitivities and convergence behaviors.
To address these challenges, we introduced a projected-PDF representation of joint-distribution information and adopted self-adaptive loss balancing at the level of statistical families. 
We also developed a reference-normalized validation framework for interpreting finite-sample statistical discrepancies and investigated the dimension of the driving white noise as a model-design choice.

The coupled-Duffing benchmark demonstrated that the projected-PDF constraints progressively improve joint-distribution accuracy and that self-adaptive loss balancing promotes more uniform accuracy across the statistical families while improving attainment of more stringent validation criteria. 
The input-dimension study further showed that increasing the driving-noise dimension can improve performance when additional independent stochastic inputs are beneficial, whereas additional noise dimensions do not necessarily provide further improvement. 
The FPU and LLV applications further demonstrated the framework on coarse observables derived from substantially higher-dimensional stochastic systems. 
For the FPU chain, multidimensional SINN reproduced the principal temporal and pairwise distributional statistics of two- and three-component observables formed from spatially separated particle positions. 
For the LLV model, SINN reproduced the coupled temporal and distributional behavior of coarse prey--predator densities obtained from microscopic KMC simulations and maintained comparable statistical agreement over the continuation segment beyond the retained window used during training.

Taken together, these results support multidimensional SINN as a practical stochastic surrogate for generating coarse vector-valued trajectories while reproducing prescribed temporal and cross-component statistics. 
At the same time, the statistical fidelity of the learned process is necessarily tied to the statistical objectives used during training. 
A finite set of projections does not uniquely determine the complete multivariate distribution, and temporal structure not adequately constrained by the training objective is not guaranteed to be reproduced.

Future work may extend the framework to higher-dimensional systems through more scalable statistical objectives and adaptive selection of projection directions. 
Additional directions include incorporating problem-specific physical constraints and extending SINN to nonstationary and parameter-dependent stochastic processes, together with longer-horizon objectives for capturing multiscale temporal dependence.

%
%


\funding{This material is based upon work supported by the U.S.\ Department of Energy (DOE), Office of Science, Office of Advanced Scientific Computing Research (ASCR), Funding for Accelerated, Inclusive Research (FAIR) program under Award Number DE-SC0024240 and under Contract Number DE-AC02-05CH11231.}



\suppdata{Supplementary material provides supporting analyses for the coupled Duffing benchmark and additional details for the FPU and LLV applications.}

\appendix

\section{Estimation of Statistical Quantities}
\label{app:est_stat}

\subsection{Standardization} 
\label{subsec:standardization} 

For a target process $\boldsymbol{X}_t\in\mathbb{R}^d$ represented in physical coordinates, the standardized process $\boldsymbol{x}_t$ is defined componentwise by
\begin{equation}
  x_{i,t} = \frac{X_{i,t}-\mu_i}{\sigma_i},
  \quad i=1,\ldots,d,
  \label{eq:standardize}
\end{equation}
where $\mu_i$ and $\sigma_i$ denote the sample mean and sample standard deviation of component $X_i$ estimated from the target ensemble.
The SINN is trained to generate trajectories $\widehat{\boldsymbol{x}}_t$ in the same standardized coordinate system, allowing statistical quantities of the target and generated ensembles to be compared directly.
Accordingly, physical-coordinate trajectories are recovered from SINN-generated trajectories by applying the corresponding inverse transformation with the same centering and scaling constants used to standardize the target process.
\begin{equation} 
    \widehat{X}_{i,t} = \mu_i + \sigma_i\, \widehat{x}_{i,t},\quad i=1,\ldots,d.
\end{equation} 
Hereafter, all statistical quantities are computed using the standardized coordinates unless stated otherwise.

\subsection{Time-Correlation Statistics} 
\label{subsec:time_correlation_statistics} 

For an ensemble of trajectories $\boldsymbol{y}=\{\boldsymbol{y}_{1:T}^{(n)}\}_{n=1}^{N}$, which may denote either the target ensemble $\boldsymbol{y}^\mathrm{tar}=\boldsymbol{x}$ or the SINN-generated ensemble $\boldsymbol{y}^\mathrm{gen}=\widehat{\boldsymbol{x}}$, the correlation function is estimated by
\begin{equation}
    C_{ij}(k) = \frac{1}{\varsigma_i \varsigma_j} \frac{1}{NT} \sum_{n=1}^N \sum_{t=1}^{T-k}
    \bigl(y_{i,t}^{(n)} - \overline{y}_i\bigr) \bigl(y_{j,t+k}^{(n)} - \overline{y}_j\bigr),
    \quad k=0,\ldots,K-1.
    \label{eq:corr_estimator}
\end{equation}
where
\begin{equation}
    \overline{y}_i = \frac{1}{NT} \sum_{n=1}^N \sum_{t=1}^{T} y_{i,t}^{(n)}, \quad
    \varsigma_i^{\,2} = \frac{1}{NT} \sum_{n=1}^N \sum_{t=1}^{T} 
    \bigl(y_{i,t}^{(n)} - \overline{y}_i\bigr)^2.
\end{equation}
The diagonal entries $C_{ii}$ correspond to the autocorrelation functions (ACFs), whereas the off-diagonal entries $C_{ij}$ ($i\neq j$) correspond to the cross-correlation functions (CCFs).
Note that the normalization factor $NT$ is used for all lag values. 
Consequently, Eq.~\eqref{eq:corr_estimator} corresponds to a biased correlation estimator containing the finite-window factor $1-k/T$, which progressively reduces the contribution of larger lag values.
Because the same estimator is applied to both the target and generated ensembles, the same finite-window weighting is applied consistently in their comparison.

We consider two variants of the correlation loss: (i) the absolute loss, defined as the mean squared error (MSE) between the target and generated correlation functions, and (ii) the relative loss, obtained by normalizing the MSE by the mean squared magnitude of the target correlation function.
For example,
\begin{equation}
    \mathcal{L}_\mathrm{ACF}^\mathrm{abs} = 
    \frac{1}{dK} \sum_{i=1}^d \sum_{k=0}^{K-1} 
    \left[ C_{ii}^{\,\mathrm{gen}}(k) - C_{ii}^{\,\mathrm{tar}}(k) \right]^2, 
    \quad
    \mathcal{L}_\mathrm{ACF}^\mathrm{rel} =
    \frac{\mathcal{L}_\mathrm{ACF}^\mathrm{abs}}{\frac{1}{dK} \sum_{i=1}^d \sum_{k=0}^{K-1} 
    \left[ C_{ii}^{\,\mathrm{tar}}(k) \right]^2}.
\end{equation}
We refer to $\mathcal{L}_\mathrm{ACF}^\mathrm{abs}$ and $\mathcal{L}_\mathrm{ACF}^\mathrm{rel}$ as the absolute and relative ACF losses, respectively.
Both absolute and relative forms are considered in this work, with the choice specified for each numerical study. 
When the corresponding population statistic is known a priori to vanish identically, as for cross-correlations between independent component processes, the absolute form is used to avoid normalization by a target statistic whose population value is zero.
The losses associated with the CCFs ($\mathcal{L}_\mathrm{CCF}^\mathrm{abs}$ and $\mathcal{L}_\mathrm{CCF}^\mathrm{rel}$) and the ACFs of the squared component processes ($\mathcal{L}_\mathrm{ACF2}^\mathrm{abs}$ and $\mathcal{L}_\mathrm{ACF2}^\mathrm{rel}$) are defined analogously.

\subsection{PDF-Based Statistics} 
\label{subsec:pdf_statistics}

Probability-density statistics are estimated using Gaussian kernel density estimation (KDE). 
For an ensemble of trajectories $\boldsymbol{y}=\{\boldsymbol{y}_{1:T}^{(n)}\}_{n=1}^{N}$, which may denote either the target ensemble $\boldsymbol{y}^{\mathrm{tar}}=\boldsymbol{x}$ or the SINN-generated ensemble $\boldsymbol{y}^{\mathrm{gen}}=\widehat{\boldsymbol{x}}$, all samples are pooled across trajectories and time points. 
Consequently, for each one-dimensional quantity of interest, a total of $M=NT$ samples $\{s_m\}_{m=1}^{M}$ are obtained from $\boldsymbol{y}$.
Examples include the marginal PDF of an individual component process and the projected PDFs introduced in Section~\ref{subsec:projected_pdf}.

For the pooled samples $\{s_m\}_{m=1}^{M}$, the KDE is defined by 
\begin{equation} 
    f(z) = \frac{1}{Mh} \sum_{m=1}^{M} \mathcal{K}\!\left( \frac{z-s_m}{h} \right), 
    \label{eq:kde1d} 
\end{equation} 
where $\mathcal{K}$ denotes the one-dimensional standard Gaussian kernel. 
The bandwidth is selected according to Scott's rule~\cite{Scott2015}. 
For one-dimensional KDEs, Scott's rule implies that the bandwidth scales as $h \propto M^{-1/5}$.

The PDF-based losses are computed as MSE discrepancies between the target and generated KDEs evaluated on a common grid with $N_z$ grid points.
As for the correlation losses, we consider absolute and relative versions of the PDF losses.
For marginal PDFs,
\begin{equation}
    \mathcal{L}_\mathrm{PDF}^\mathrm{abs} = \frac{1}{d N_z} \sum_{i=1}^{d} \sum_{r=1}^{N_z}
    \left[ f_i^{\,\mathrm{gen}}(z_r) - f_i^{\,\mathrm{tar}}(z_r) \right]^2,
    \quad
    \mathcal{L}_\mathrm{PDF}^\mathrm{rel} = \frac{\mathcal{L}_\mathrm{PDF}^\mathrm{abs}}{\frac{1}{d N_z} \sum_{i=1}^{d} \sum_{r=1}^{N_z} \left[ f_i^{\,\mathrm{tar}}(z_r) \right]^2}.
\end{equation}
For projected PDFs,
\begin{equation}
\begin{split}
    \mathcal{L}_\mathrm{PROJ}^\mathrm{abs} &= \frac{1}{N_\mathrm{proj}N_z} 
    \sum_{\boldsymbol{w}\in\mathcal{W}_\mathrm{proj}} \sum_{r=1}^{N_z}
    \left[ f_{\boldsymbol{w}}^{\,\mathrm{gen}}(z_r) - f_{\boldsymbol{w}}^{\,\mathrm{tar}}(z_r) \right]^2,
    \\
    \mathcal{L}_\mathrm{PROJ}^\mathrm{rel} &= \frac{\mathcal{L}_\mathrm{PROJ}^\mathrm{abs}}{\frac{1}{N_\mathrm{proj}N_z} \sum_{\boldsymbol{w}\in\mathcal{W}_\mathrm{proj}} \sum_{r=1}^{N_z}
    \left[ f_{\boldsymbol{w}}^{\,\mathrm{tar}}(z_r) \right]^2}.
\end{split}
\end{equation}

\section{Construction of Projection Directions: Rank and Conditioning Analysis} 
\label{app:projection_directions}

The projected-PDF methodology introduced in Section~\ref{subsec:projected_pdf} represents joint-distribution information through a finite collection of one-dimensional projections. 
Because only finitely many projection directions are employed in practice, an important question is how these directions should be selected. 
In this appendix, we describe the linear-algebraic considerations that guided the construction of the projection directions used in this work.

\subsection{Cumulant coefficient matrices and rank criterion} 

As discussed in Section~\ref{subsec:projected_pdf}, the cumulants of a projected variable are related to the joint cumulants of the underlying random vector through Eq.~\eqref{eq:cumulant_multilinearity}.
The number of distinct joint cumulants of total order $m$ in $d$ dimensions is 
\begin{equation} 
    M_d(m) = \binom{m+d-1}{d-1}. 
    \label{eq:appendix_cumulant_count} 
\end{equation} 
Accordingly, the vector of order-$m$ joint cumulants may be represented by a vector $\boldsymbol{\kappa}^\mathrm{joint}_m \in \mathbb{R}^{M_d(m)}$. 
For a projection direction $\boldsymbol{w}$ and cumulant order $m$, let 
\begin{equation} 
    \mathbf{c}_m(\boldsymbol{w}) 
    = \left[ \frac{m!}{\alpha!}\, \boldsymbol{w}^{\alpha} \right]_{|\alpha|=m} 
\end{equation} 
denote the row vector containing the coefficients appearing in Eq.~\eqref{eq:cumulant_multilinearity}. 
Since there are $M_d(m)$ multi-indices satisfying $|\alpha|=m$, the vector $\mathbf{c}_m(\boldsymbol{w})$ contains $M_d(m)$ components. 
Stacking the coefficient vectors associated with all projection directions yields the coefficient matrix 
\begin{equation} 
    \mathbf{C}_m = 
    \begin{bmatrix} 
        \mathbf{c}_m(\boldsymbol{w}_1) \\ 
        \mathbf{c}_m(\boldsymbol{w}_2) \\ 
        \vdots \\
        \mathbf{c}_m(\boldsymbol{w}_{N_{\mathrm{dir}}}) 
    \end{bmatrix} 
    \in \mathbb{R}^{N_{\mathrm{dir}}\times M_d(m)} . 
\end{equation} 
Let $\boldsymbol{\kappa}^{\mathrm{proj}}_m$ denote the vector containing the order-$m$ cumulants of the projected variables. 
Equation~\eqref{eq:cumulant_multilinearity} may then be written compactly as 
\begin{equation} 
    \boldsymbol{\kappa}^{\mathrm{proj}}_m = \mathbf{C}_m \boldsymbol{\kappa}^\mathrm{joint}_m .
    \label{eq:linear_cumulant_relation} 
\end{equation}
The coefficient matrix $\mathbf{C}_m$ therefore characterizes how the selected projection directions constrain the order-$m$ joint cumulants.

The order-$m$ joint cumulants can be uniquely determined from the projected cumulants if and only if
\begin{equation} 
    \operatorname{rank}(\mathbf C_m) = M_d(m). 
\end{equation}
Accordingly, we selected projection directions by seeking collections that satisfy this rank condition for as many cumulant orders as possible while keeping the total number of directions moderate and maintaining favorable conditioning.

\subsection{Rank analysis of the selected direction sets} 

For the two-dimensional case, the complete direction set consists of the two coordinate directions together with the six non-coordinate directions generated by the three orthonormal frames introduced in Eq.~\eqref{eq:frames2d}. 
The resulting collection contains $N_{\mathrm{dir}} = 8$ projection directions. 
The associated coefficient matrices $\mathbf{C}^{(2)}_m$ were constructed from Eq.~\eqref{eq:cumulant_multilinearity} using these eight directions.
Direct evaluation yields 
\begin{equation} 
    \operatorname{rank} \!\left( \mathbf{C}^{(2)}_m \right) = M_2(m), \qquad m=1,\ldots,7 .
\end{equation} 
Thus, all two-dimensional joint cumulants through order seven are uniquely identifiable from the selected projections.
The first rank deficiency occurs at order eight: 
\begin{equation} 
    \operatorname{rank} \!\left( \mathbf{C}^{(2)}_8 \right) = 8 < M_2(8) = 9 . 
\end{equation} 
Consequently, the eighth-order joint cumulants cannot be uniquely determined from this direction set.
We note that the selected eight directions achieve full column rank for every cumulant order for which full column rank is theoretically possible. 
Since $M_2(m)=m+1$ and $\mathbf{C}^{(2)}_m$ has only eight rows, full column rank cannot be achieved for $m\ge 8$.

The same analysis was performed for the three-dimensional direction set introduced in Eq.~\eqref{eq:frames3d}, which consists of three coordinate directions and nine non-coordinate directions for a total of $N_{\mathrm{dir}}=12$ directions. 
The corresponding coefficient matrices satisfy 
\begin{equation} 
    \operatorname{rank} \!\left( \mathbf{C}^{(3)}_m \right) = M_3(m), \qquad m=1,2,3 , 
\end{equation} 
whereas 
\begin{equation} 
    \operatorname{rank} \!\left( \mathbf{C}^{(3)}_4 \right) = 12 < M_3(4) = 15 .
\end{equation} 
Thus, all three-dimensional joint cumulants through order three are uniquely identifiable from the selected projections.
As in the two-dimensional case, full column rank is achieved for every cumulant order for which full column rank is theoretically possible, namely all orders satisfying $M_3(m)\le 12$.

\subsection{Conditioning analysis} 

Full column rank alone does not guarantee that the projection directions provide numerically useful constraints. 
If the columns of $\mathbf{C}_m$ are nearly linearly dependent, small errors in estimated projected cumulants may be amplified substantially when inferring information about the joint cumulants. 
To assess numerical robustness, we therefore examined the spectral condition number $\kappa_2(\mathbf{C}_m) = \|\mathbf{C}_m\|_2 \left\| \mathbf{C}_m^{\dagger} \right\|_2$,
where $\mathbf{C}_m^{\dagger}$ denotes the Moore--Penrose pseudoinverse. 
Smaller condition numbers indicate more stable identification of the joint cumulants from the projected cumulants.

For the particular direction sets defined in Eqs.~\eqref{eq:frames2d} and~\eqref{eq:frames3d},
\begin{equation} 
    \max_{1\le m\le7} \kappa_2 \!\left( \mathbf{C}^{(2)}_m \right) = 7.951, 
\end{equation} 
whereas
\begin{equation} 
    \max_{1\le m\le3} \kappa_2 \!\left( \mathbf{C}^{(3)}_m \right) = 4.410. 
\end{equation}
These moderate condition numbers indicate favorable numerical conditioning for identifying the joint cumulants from the selected projections.
Note, however, that the condition numbers reported here are computed using the unnormalized unit-vector projections appearing in Eq.~\eqref{eq:cumulant_multilinearity}.
The normalization in Eq.~\eqref{eq:projected_variable} preserves the rank of the coefficient matrices but may change their condition numbers through direction-dependent row scaling.

The rank and conditioning analyses presented above characterize, respectively, the identifiability of low-order joint cumulants from the selected projections and the numerical stability of that identification; they do not imply recovery of the complete multivariate distribution from a finite set of projections.


\newpage

\includepdf[pages=-, width=\textwidth, pagecommand={}]{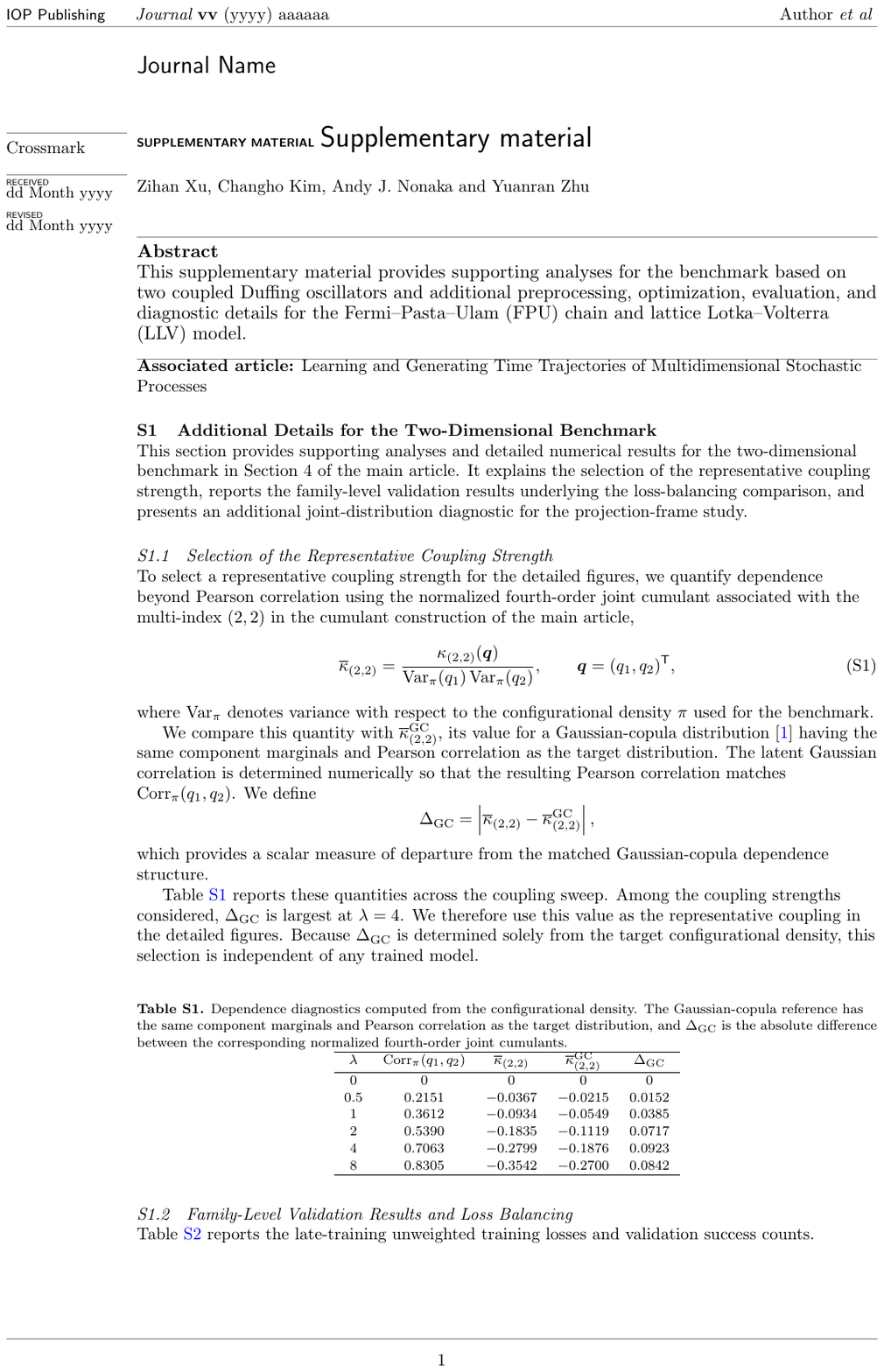}

\end{document}